\documentclass{emulateapj}
\usepackage{epsfig}
\usepackage{amsmath}
\usepackage{amssymb}
\usepackage{natbib}
\usepackage{setspace}
\usepackage{graphicx}
  \usepackage[flushleft]{threeparttable}

\begin{document}
\title{A model for the onset of self-gravitation and star formation in molecular gas governed by galactic forces: I. cloud-scale gas motions}
\author{Sharon E. Meidt\altaffilmark{1}}
\author{Adam K. Leroy\altaffilmark{2}}
\author{Erik Rosolowsky\altaffilmark{3}}
\author{J.~M.~Diederik Kruijssen\altaffilmark{4}}
\author{Eva Schinnerer\altaffilmark{1}}
\author{Andreas Schruba\altaffilmark{5}}
\author{Jerome Pety\altaffilmark{6,7}}
\author{Guillermo Blanc\altaffilmark{8,9,10}}
\author{Frank Bigiel\altaffilmark{11}}
\author{Melanie Chevance\altaffilmark{4}}
\author{Annie Hughes\altaffilmark{12,13}}
\author{Miguel Querejeta\altaffilmark{14}}
\author{Antonio Usero\altaffilmark{15}}

\altaffiltext{1}{Max-Planck-Institut f\"ur Astronomie / K\"{o}nigstuhl 17 D-69117 Heidelberg, Germany}
\altaffiltext{2}{Department of Astronomy, The Ohio State University, 140 W. 18th Ave., Columbus, OH 43210, USA} 
\altaffiltext{3}{Department of Physics, University of Alberta, Edmonton, AB, Canada} 
\altaffiltext{4}{Astronomisches Rechen-Institut, Zentrum f\"{u}r Astronomie der Universit\"{a}t Heidelberg, M\"{o}nchhofstra\ss e 12-14, 69120 Heidelberg, Germany} 
\altaffiltext{5}{Max-Planck-Institut ff\"ur extraterrestrische Physik, Giessen-bachstrasse 1, 85748 Garching, Germany} 
\altaffiltext{6}{Institut de Radioastronomie Millim\'etrique, 300 Rue de la Piscine, F-38406 Saint Martin d'H\`eres, France}
\altaffiltext{7}{LERMA, Observatoire de Paris, PSL Research University, CNRS, Sorbonne
  Universit\'es, UPMC Univ. Paris 06, \'Ecole normale supérieure, F-75005, Paris, France}
\altaffiltext{8}{Departamento de Astronomia, Universidad de Chile, Casilla 36-D, Santiago, Chile}
\altaffiltext{9}{Centro de Astrofisica y Tecnologias Afines (CATA), Camino del Observatorio 1515, Las Condes, Santiago, Chile}
\altaffiltext{10}{Visiting Astronomer, Observatories of the Carnegie Insti- tution for Science, 813 Santa Barbara St, Pasadena, CA, 91101, USA}
\altaffiltext{11}{Institut f\"{u}r theoretische Astrophysik, Zentrum f\"{u}r Astronomie der Universit\"{a}t Heidelberg, Albert-Ueberle Str. 2, D-69120 Heidelberg, Germany}
\altaffiltext{12}{CNRS, IRAP, 9 av. du Colonel Roche, BP 44346, F-31028 Toulouse cedex 4, France}
\altaffiltext{13}{Universite de Toulouse, UPS-OMP, IRAP, F-31028 Toulouse cedex 4, France}
\altaffiltext{14}{European Southern Observatory, Karl-Schwarzschild-Str. 2, D-85748 Garching, Germany}
\altaffiltext{15}{Observatorio Astron\'{o}mico Nacional - OAN, Observatorio de Madrid Alfonso XII, 3, 28014 - Madrid, Spain}

\date{\today}
\begin{abstract}
Modern extragalactic molecular gas surveys now reach the scales of star-forming giant molecular clouds (GMCs, 20-50 pc). Systematic variations in GMC properties with galaxy environment imply that clouds are not universally self-gravitating objects, decoupled from their surroundings.  Here we reexamine the coupling of clouds to their environment and develop a model for 3D gas motions generated by forces arising with the galaxy gravitational potential defined by the background disk of stars and dark matter.  We show that these motions can resemble or even exceed the motions needed to support gas against its own self-gravity throughout typical galaxy disks.  The importance of the galactic potential in spiral arms and galaxy centers suggests that the response to self-gravity does not always dominate the motions of gas at GMC scales, with implications for observed gas kinematics, virial equilibrium and cloud morphology.  We describe how a uniform treatment of gas motions in the plane and in the vertical direction synthesizes the two main mechanisms proposed to regulate star formation: vertical pressure equilibrium and shear/Coriolis forces as parameterized by Toomre Q$\approx$1. As the modeled motions are coherent and continually driven by the external potential, they represent support for the gas that is distinct from that conventionally attributed to turbulence, which decays rapidly and requires thus maintenance, e.g. via feedback from star formation.  Thus our model suggests the galaxy itself can impose an important limit to star formation, as we explore in a second paper in this series. 
\end{abstract}

\section{Introduction\label{sec:intro}}

The origin of observed motions in the dense star-forming phase of the ISM is a topic undergoing renewed debate, with implications for the nature of turbulence, the regulation of star formation, and the rate at which galaxies convert their cold gas in to stars. In this paper, we explore the possibility that turbulent motions originate as coherent, orbital motions driven by the gravity associated with the host galaxy. This should produce a clear, observable dependence of molecular cloud properties on galactic environment, and will introduce an observable dependence of star formation rate per unit gas on galactic kinematics (S. Meidt et al. in preparation).

Conventionally, the super-thermal line widths exhibited by emission-line tracers of the cold ISM are interpreted as the signature of turbulent motions. These have been observed to have magnitude that approximates the self-gravity of the gas.  This ability of turbulence to support star-forming gas clouds is thought to help explain the inefficiency of star formation in the cold ISM. Without support from turbulent motions, gas would form stars much more rapidly than observed \citep{zuck}.

Observed gas motions near the scales at which gas becomes self-gravitating are thought to reflect the processes by which turbulence originates and decays. The relation between the line-widths and the sizes of molecular clouds has been argued to be a manifestation of the characteristic cascade of energies associated with super-sonic turbulence (\citealt{larson}; \citealt{elmscalo}; \citealt{mckeeOst}; \citealt{kritsuk}). Simulations that drive turbulence through SNe explosions \citep{padoan15} or via gas accretion on to the galactic disk \citep{gnedin} are both able to reproduce the observed relation between cloud motions and size.

In other simulations, the linewidth-size relation emerges entirely as a consequence of gas self-gravity (\citealt{ibanez}; \citealt{camacho}). This supports an earlier proposal that the motions within a cloud arise as a byproduct of gravitational collapse (\citealt{vs08}; \citealt{ballesteros}).  In this scenario, observed internal turbulent motions resemble a state of energy equipartition rather than a supported cloud in equilibrium.

Clouds are observed to  an additional dependence of between velocity dispersion and cloud surface density \citep{heyer} at a given size scale (formally $\Sigma \propto \sigma^2/R$). This dependence has been seen in extragalactic molecular gas surveys that collectively achieve a high dynamic range in cloud-scale surface density (\citealt{hughesI}; \citealt{leroy2015};\citealt{leroy2016}; J. Sun et al. in preparation). This relationship has been interpreted to indicate proximity to virialization or energy equipartition, with kinetic energy about equal to potential energy. Cloud-scale motions originating with gravity tend to be consistent with such observations because collapse motions would naturally match the gas self-gravity.

Despite this overall scaling, observed clouds do commonly exhibit 'super-virial' motions, showing systematic departures from the above relation. These occur in the MW center \citep{oka,kruijssen14}, in some starburst galaxies (e.g., \citealt{Johnson15}. They appear common in the outer disk of the Milky Way (e.g., \citealt{heyer}) and the atomic-rich, low surface density disks of the Local Group spirals M31 and M33 (\citealt{leroy2016}, J. Sun et al. in preparation).

So far, the most popular explanation for these super-virial motions is a confining external pressure. Such pressure, acting on a cloud boundary, will raise the line width of a cloud with a particular size and surface density (\citealt{elm93}; \citealt{field} and see \citealt{hughesI}).  Preliminary comparisons between internal cloud pressure and the hydrostatic mid-plane pressure suggest that external pressure may be sufficient to play such a role in normal disk galaxies, particularly those dominated by atomic gas (\citealt{hughesI}, \citealt{hughesIII}; A. Schruba et al. in preparation).

So far, most studies have considered ISM pressure of indeterminate origin, focusing mainly on the existence of deviations from the pure self-gravitating case. The physical origin of the environmental contribution to cloud line widths and dynamical state will have important implications. If some environments preferentially support clouds against collapse, we can expect molecular cloud populations and the star-forming properties of the gas to vary in response. We are entering the era of large surveys of giant molecular cloud (GMC) populations spanning diverse environments, and such variations should be observable in current and upcoming surveys.

Here, we hypothesize that some turbulent motions originate as coherent, orbital motions driven by the gravity associated with the host galaxy. Depending on the local galactic potential, these motions may dominate those induced by self-gravity, leading to the observed deviations from self-gravity described above.

The driving of turbulence in relation to large scale kinematics that prompt instabilities in the gas has been widely recognized. Previous studies have highlighted the importance of galactic rotation \citep{fleck}, magneto-rotational instabilities (\citealt{sbalbus}; \citealt{kim03}) and swing-amplified shear instabilities (\citealt{Gammie01}; \citealt{huber}; \citealt{wada}). Spiral arms can are thought to provide an important avenue for generating turbulence \citep{elm03}, for example via the Raleigh-Taylor instabilities that also lead to spur formation \citep{kko06} or the passage of clumpy gas through the spiral shock (\citealt{bonnell}; \citealt{dobbs07}). In some galaxies, strong non-circular motions have been observed, which should lead to converging flows and strong shear.  These conditions have also been suggested to create turbulence via non-linear thin-layer phase instability in converging flows (\citealt{vs06}; \citealt{vishniac}; \citealt{heitsch05}; \citealt{heitsch06}).   

Here, we consider a gas disk in which fluid elements populate nearly-circular orbits determined by the shape of the gravitational potential of the galaxy. We examine how departures from circular motion create coherent motions on the cloud scale. The strength of these motions depend on the shape of the rotation curve and so reflect a galaxy's large scale potential, and are directly observable. From here we introduce a model for gas motions at the cloud scale that combines the effects of gas self-gravity and motions driven in response to forces exerted by the background host galaxy ($\S$ \ref{sec:themodel}). The stellar and dark matter distributions, in which the gas is embedded, define an external potential that acts on the gas.  After reconsidering the forces relevant for internal cloud motions, in particular Coriolis forces that arise due to the motion of clouds about the galaxy center ($\S$ \ref{sec:forces}), we model the resulting motions in $\S$ \ref{sec:epicycles}. Then we compare relative strengths of self-gravity and the background potential in $\S$ \ref{sec:quantplots}.  

As a result of the properties (mass, scale length, scale height) of typical galaxy disks, gravitationally-induced motions contribute to systematic variations in the dynamical state of the gas with location in the galaxy (center, disk, outskirts), as we explore in $\S$ \ref{sec:dynstate}.  We undertake a preliminary comparison of the model to observations of molecular gas on cloud scales in $\S$ \ref{sec:datacomp}.  Then in $\S$ \ref{sec:predictions} we describe several other predictions of the model that can be tested with observations, including implications for virial equilibrium and variations in the CO-to-$H_2$ conversion factor.  We summarize and conclude in $\S$ \ref{sec:summary}.

This work complements recent work that considers turbulent pressure originating with star formation feedback or hierarchical collapse under the influence of self-gravity (i.e. \citealt{krumBurk17}, or with processes occurring beyond the cloud edge (at the outer scale of turbulence; i.e. \citealt{krumKruijs}; \citealt{krumholz2017}). As we lay out below, our proposal can be tested and compared to these other hypotheses by combining kinematic measurements with studies of resolved GMC populations.

In the second paper of this series, we study the implications of this new model for the process of star formation throughout galaxies.  Self-gravity must overcome the motions defined by the background galaxy potential in order to collapse and form stars.  Thus our model signifies a path toward inefficient star formation regulated by the galaxy disk itself, as will be discussed in paper~II.  

\section{The model}
\label{sec:themodel}

In this section we introduce a model for motions within molecular gas due to the background rotating distribution of stars and dark matter.  The gravitational potential defined by the background host galaxy exerts forces that drive epicyclic motions that are still significant on the scale of individual molecular clouds.

These motions can be isotropic or non-isotropic in three dimensions, depending on the scale of the cloud relative to the stellar and dark matter mass distribution. For a given spatial scale, $R_c$, we calculate these motions to be

\begin{equation}
\begin{cases}
3\sigma_{gal, iso}^2=3(\kappa R_c)^2 \hspace*{1.5cm}\text{isotropic motions}\\
3\sigma_{gal}^2=2(\kappa R_c)^2+(\nu R_c)^2 \hspace*{.1cm}\text{non-isotropic motions}~.
\end{cases}
\label{eq:siggalcases}
\end{equation}

Here $\kappa$ is the frequency of radial oscillations about the (roughly) circular orbit traced by the cloud about the galaxy center and $\nu$ is the frequency of vertical oscillations about the galactic mid-plane. Both frequencies, $\kappa$ and $\nu$, are determined by the distribution of stellar and dark matter mass in the galaxy.  The relevant size scale $R_c$ for the present study is typically the size of a cloud, so these quantities indicate the motions induced by the galactic potential on cloud scales.

In-plane motions due to the differential rotation in the host galaxy potential appear in both cases (in the isotropic case and in the first term of the second, non-isotropic case).  We emphasize that as long as velocity field of the galaxy exhibits coherent behavior down to $R_c$, these motions {\em must} exist.  Motions perpendicular to the disk of the galaxy, which are also included in both cases, must also exist to balance the potential of the galaxy over the scale of the cloud.  Depending on the vertical extent of the cloud, these can be the same as in-plane motions (isotropic case) or deviate from the in-plane motions (non-isotropic case).

The isotropic case should be relevant when $R_c$ is small relative to the scales over which the background density changes.  The background potential in this scenario approximates the potential of a uniform density field, so that the motions induced by this field look the same in all three dimensions.

As we show in the next section, under typical conditions in nearby galaxies the motions arising with the galaxy potential can be comparable to the motions associated with self-gravity on cloud scales. These motions also have velocities with the same magnitude as observed gas velocity dispersions. Including these motions in a model of the cloud dynamical state yields a different picture than the conventional view of decoupled, self-gravitating clouds.  

The observational signatures implied by our calculations are discussed in $\S$ \ref{sec:quant}, where we also demonstrate consistency with several recent surveys of GMCs. In $\S$ \ref{sec:predictions} we comment on additional signatures of the influence of the galactic potential on cloud scales that could be tested in future analysis. 

Throughout, self-gravitation refers to the particular state in which the potential of the gas (the gas 'self-gravity') dominates over any other (gravitational or magnetic) potential present.  In this work we focus on the host galaxy gravitational potential and thus gas is called `self-gravitating' when the motions due to 'self-gravity' would exceed the motions due to the galaxy potential. This terminology deviates somewhat from the definition of self-gravitation in more wide use, in which motions due to self-gravity dominate over turbulent motions.  Assessing self-gravitation by that measure requires a model for scale-dependent turbulence distinct from other gravitationally-induced motions, and is beyond the scope of this work.

\subsection{The basic framework}

We envision clouds as coherent structures in position and velocity space with smoothly varying density distributions. These are not necessarily gravitationally bound objects with well-defined borders, and the density may vary smoothly across the edge of a cloud into the surrounding medium. For our purpose, the size of a cloud will be set by the scale over which the combined gravitational potentials of the stars, dark matter, and the gas itself act coherently on the gas.

In practice, we expect this size scale to resemble the measured sizes of GMCs.  Thus we label parcels of gas as `clouds' and, when we refer to other observable properties such as mass and surface density, we take these to be measured at a size scale $R_c$. 
In Paper II (S. Meidt et al. in preparation), we identify a more physical cloud radius, the scale at which gravitational forces balance. 

The gas that makes up a cloud is assumed to be isothermal and inviscid.  Although turbulence is implicitly assumed to be present, the motions of the gas that we are primarily concerned with are gravitational in origin.  We also neglect magnetic forces, although we note that these are likely to be important for organizing and structuring the gas on galactic scales (i.e. \citealt{ko02}; \citealt{ko06}).  

The cloud material is further assumed to belong to a rotating gas disk within which motions are established by the current distribution of orbital energies. 
This distribution is implicitly taken to have evolved from the initial energy distribution inherited at the time of accretion (followed by early cooling to a cold, neutral phase) and all subsequent accretion events (e.g. \citealt{elmBurk}; \citealt{krumBurk}).  As the gas settles into a cold disk, gravitational (and viscous) torques continually move energy inward (i.e. \citealt{krumKruijs}; \citealt{KimElmegreen2017}), while collisions and shocks in the cold gas result in dissipation, eventually leading to a lowest-energy configuration of circular orbits.  Until that time, the distribution of orbital energies establishes a gas disk in which fluid elements populate nearly-circular orbits determined by the shape of the gravitational potential of the galaxy.  As described in detail below, these departures from circular motion constitute a field of coherent motions on the cloud scale, given the typical observed characteristics of galaxy disks (shape and rotational velocities).   

\subsection{Inventory of relevant forces and motions}
\label{sec:forces}

Consider a cloud at galactocentric radius $R_{gal}$ and density $\rho$ rotating about the center of a galaxy with angular speed $\Omega$.  In the rotating frame of reference in which the cloud's center of mass is stationary, the Euler equation of motion for a fluid element within the cloud is
\begin{equation}
\frac{\partial\vec{v}}{\partial t}=-\nabla\Phi_c-\nabla\Phi_{gal}-\nabla h-\vec{\Omega} \times (\vec{\Omega}\times\vec{r})-2\vec{\Omega} \times \vec{v} \label{eq:Euler}
\end{equation}  
Here $\Phi_c$ is the potential of the cloud material and $\Phi_{gal}$ is the galactic potential defined by the distribution of stars and dark matter. 

The last two terms on the right are the fictitious Centrifugal and Coriolis forces in the rotating frame of reference, respectively. The third term involving the specific enthalpy $h$ represents the pressure gradient in the gas, i.e., $dh=dP/\rho$.  This factor is neglected in the case of a collisionless system (stars) or when the density is everywhere uniform\footnote{As for a collisionless stellar system, when applied to a system of ballistic clouds (as opposed to an individual fluid element within clouds, as here), the Euler equation of motion is typically written without the third, pressure gradient term.}. We ignore additional forces, e.g., due to magnetic fields, as we are focused on the effect of the galaxy's potential in this paper.   

The centrifugal force, together with the force exerted by the galaxy, is responsible for the tidal force experienced by clouds, 
\begin{equation}
F_{tide}=\nabla\Phi_{gal}+\Omega^2 r~.
\end{equation}
This tidal force elongates or even destroys self-gravitating objects (i.e., \citealt{stark}; \citealt{das}; \citealt{renaud}; \citealt{ballesteros09}). This is the most typical form in which the impact of the host galaxy on clouds has been considered. 

A handful of extragalactic GMC studies have measured the stability of clouds against deformation or destruction by tides (\citealt{stark}; \citealt{blit85}; \citealt{ros05}; \citealt{thilliez}). These treatments have adopted the limit of instantaneous relaxation, effectively assuming that all parts of the object follow the same circular orbit, ignoring the potentially important role of the Coriolis force. Such an approximation is valid as long as the Coriolis force due to internal motions (or by other forces present in the object that are ignored here, i.e. magnetic fields in clouds; \citealt{mestelMag}; \citealt{habe}; \citealt{Mouschovias}).   

Ignoring the Coriolis force should be valid in the case of star clusters, which have higher densities than GMCs by more than an order of magnitude. But this is not clear for typical giant molecular clouds.  We show below that on GMC scales, the motions that give rise to the Coriolis force can have the same magnitude as the motions associated with the cloud's self-gravity. This effect will be strongest for large clouds in the centers of galaxies and at the locations of strong perturbations to the potential, like spirals arms and bars.

In this treatment, we include the influence of the Coriolis force. We compare the motions that arise due to the potential of the host galaxy to those that should arise due only to the cloud's own weight.

\subsection{Epicyclic motion}\label{sec:epicycles}

First, we consider the limit of no gas self-gravity and uniform density.  As a result of the collisional, dissipational nature of gas, we assume that each parcel of gas follows a nearly circular orbit (e.g., \citealt{gammie}). A cloud represents a collection of such objects, each executing their own, slightly different orbit.  In this picture, a cloud is not a ballistic object like a single star, but rather a collection of elements that move independently, like a set of independent neighboring stars.  The motions of  fluid elements within the cloud can each be described as the circular motion of a guiding center along a perfectly circular orbit plus a small epicyclic excursion about the guiding center.  

The departure of an individual fluid element 
from the circular orbit traced by its guiding center is determined by the shape of the local galactic potential. In the plane of the galaxy, the fluid element traverses the well-known radial epicycle with frequency $\kappa$ (see below).  Perpendicular to the plane, the motion of an element needed to balance the potential of the galaxy can be described as vertical epicycle with frequency $\nu$.

Before proceeding with a detailed description of these epicycles, here we note the sense of the forces at play. In the rotating frame of reference of the cloud, the difference in the galactic potential across the cloud leads to a tidal force that acts along the gradient of the potential. This will usually be the parallel to the direction of the galactic center. Additionally, fluid elements farther from the galactic center move with an angular speed slower than the cloud center. They appear to travel opposite the direction of rotation in the frame of the cloud. Elements closer to the galactic center do the reverse, preceding the cloud center. As a result, the Coriolis force contributes an additional force in the direction perpendicular to the motion of the cloud and parallel to the direction of the galactic center. The overall motion of a fluid element follows that of an ellipse in a sense that is retrograde of the orbital motion about the galaxy center.

The epicyclic motion described below follows the standard description given by Binney \& Tremaine (1987).  Because the collisional nature of gas will lead to the loss of energy and angular momentum and act to depopulate all orbits except those that are not intersecting, our treatment assumes an additional conceptual simplification.  We assume that clouds are composed of fluid elements that  orbit around the galaxy center with the same angular momentum, forming a coherent unit. The motions of individual elements within the cloud are thus deconstructed into a fixed circular orbit traced by the cloud's center of mass (i.e. a single guiding center) and an epicyclic excursion about this circular orbit.  As a result, a cloud in this treatment is constructed of fully populated nested epicycles, minimizing orbital intersections.  

For the small displacements we are interested in here (see $\S$\ref{sec:epicSizes}), rms velocities across clouds in this scenario are approximately the same to lowest order as those derived with a more generic treatment.  The velocity dispersion in any region can alternatively be envisioned as reflecting the overlap of epicycles around unique, neighboring guiding centers (each associated with its own angular momentum; see Binney \& Tremaine 1987).  
Later in $\S$ \ref{sec:predictions} we describe several observational signatures of nested epicycles that can be used to test whether clouds are indeed constructed in this manner. 

\subsubsection{Paths of fluid elements}\label{sec:epicTraj}

We work in the non-inertial reference frame of the cloud's center of mass. We orient the frame with $x$-axis in the direction of the galactic center, the $y$-axis along the direction of local tangential motion, and the $z$-axis pointing in the vertical direction aligned with the angular momentum of the disk. The origin of the coordinate system is selected to sit at the cloud's center of mass at position ($x_c$, $y_c$, $z_c$) with $z_c = 0$ taken for convenience (but see below). We refer to the corresponding inertial frame with coordinates $X$, $Y$, and $Z$ with $X$ and $Y$ in the plane of the disk.

We assume that the galaxy's gravitational potential is separable into vertical and planar components, 
\begin{equation}
\Phi(x,y,z)=\Phi(x,y)+\Phi(z)\label{eq:galpot}~,
\end{equation}
and consider motions in the plane separately from those in the vertical direction. This treatment should be appropriate for disk galaxies on the scales of giant molecular clouds, and we justify it below. \\

\noindent{\underline{In the plane}}\\

In the inertial frame of reference oriented with the plane of the host galaxy disk, we write the effective potential at galactocentric radius $R$=$(X^2+Y^2)^{1/2}$ and the mid-plane of the disk ($Z = z = 0$) as 
\begin{equation}
\Phi_{eff}(X,Y)=\Phi(X,Y)+\Omega^2 R^2/2~.
\end{equation} 
\noindent For a cloud tracing a perfectly circular orbit in the inertial reference frame, the effective potential is defined to be zero at the cloud's center of mass $(x_c, y_c, z_c)$.

At all other positions in the cloud, the motions of fluid elements can be calculated by expanding the local potential around the cloud's center in the non-inertial reference frame,
\begin{eqnarray}
&\Phi_{eff}&(x,y)=\\
&\Phi_{eff}&(x_c,y_c)+\frac{\partial^2\Phi_{eff}}{\partial x^2}(x-x_c)^2+\frac{\partial^2\Phi_{eff}}{\partial y^2}(y-y_c)^2+...\nonumber
\end{eqnarray}
ignoring higher order terms and those of order $xy$ in the epicyclic approximation.  

The equations of motion  in this reference frame can be written
\begin{equation}
\frac{\partial v_x}{\partial t}=-\kappa^2 x-2\Omega v_y \label{eq:diffeqx}
\end{equation}  
and
\begin{equation}
\frac{\partial v_y}{\partial t}=2\Omega v_{x}\label{eq:diffeqy}
\end{equation}  
where 
\begin{equation}
\kappa^2=\frac{\partial^2\Phi_{eff}}{\partial x^2} \nonumber \\
\end{equation}
and the variation in the potential in the $y$-direction is assumed to be negligible compared to the $x$ direction. 

In axisymmetric potentials, the epicyclic frequency is commonly written as 
\begin{eqnarray}
\kappa^2&=&4\Omega^2+R\frac{d\Omega^2}{dR}  \nonumber \\
&=&-4B(A-B)\label{eq:kappafromomega}
\end{eqnarray}
in terms of the Oort A and B constants, which measure shear and vorticity.  

This set of differential equations (\ref{eq:diffeqx} and \ref{eq:diffeqy}) has the familiar solutions 
\begin{equation}
x=X_0 \cos{(\kappa t)} \label{eq:xmotion}
\end{equation}
and
\begin{equation}
y=\frac{2\Omega}{\kappa}X_0 \sin{(\kappa t)}\label{eq:ymotion}~.
\end{equation}
Here $X_0$ is the size of the epicycle, described more below. 

Thus in this limit, considering only the galactic potential, the fluid elements follow elliptical paths with axis ratio $2\Omega/\kappa$ in the non-inertial frame. In the case that all cloud material orbits around the galaxy with the same angular momentum, this leads to coherent, nested epicycles with a uniform shape centered on the cloud center at position ($x_c$, $y_c$, $z_c$).\\

{\noindent\underline{In the vertical direction}}\\

We describe vertical motions in a similar way to those in the plane. 
Letting the gas disk now have a vertical extent $Z_0$, we envision the gas as populating a set of roughly circular orbits that are not restricted to the mid-plane, due to a slight offset between each individual orbit's axis of rotation and the $z$-axis aligned with the net disk angular momentum.  
In the epicyclic approximation, these orbits are described as a circular part limited to $z=0$ plus a vertical excursion about this circular orbit.    

Expanding the local potential around the cloud center ($x_c$, $y_c$, $z_c$=0), the equation of motion for small vertical displacements is written
\begin{equation}
\frac{\partial v_z}{\partial t}=-\nu^2 z
\end{equation}
where 
\begin{equation}
\nu^2=\frac{\partial^2\Phi(z)}{\partial z^2}.  
\end{equation}
The solution to this equation of motion 
\begin{equation}
z=Z_0 \cos{(\nu t)} \label{eq:zmotion}
\end{equation}
represents periodic paths of fluid elements in the vertical direction.  

The motions in this epicyclic description are equivalent to those familiar from hydrostatic equilibrium in the limit of no self-gravity. When equilibrium is dominated by the gravity of the background stellar (and dark matter) distribution with density $\rho$, the gas velocity dispersion is written
\begin{equation}
\sigma^2=Z_0^24\pi G\rho.
\end{equation} 
This can be approximated as $Z_0^2\nu^2$ according to Poisson's equation, which relates the distribution of (stellar and dark matter) material to the background potential, 
\begin{equation}
4\pi G\rho\approx\frac{\partial^2\Phi_{eff}}{\partial z^2}=\nu^2. \label{eq:vertfreq1}
\end{equation}
This applies in the limit $\nu$$>$$\kappa$ applicable for the thin star-forming disks we are interested in (i.e. larger variation in the potential in the vertical direction than in the plane; see more below) and most accurately when the disk density is approximately constant (also see below).  

At the mid-plane of the disk where the molecular gas resides, the stellar contribution to the potential gradient usually dominates that associated with dark matter.\footnote{As the stellar density decreases at large galactocentric radius, the dark matter is expected to make an increasingly important contribution to the mid-plane potential.}  Assuming that the stellar disk is in equilibrium, with vertical distribution $\rho(z)=\rho_0 \exp{(-z/z_0)}$ and scale height $z_0$, we can write 
\begin{equation}
\nu^2=2\pi G\Sigma_{stars}z_0^{-1}\label{eq:vertfreq}~.
\end{equation}
$\Sigma_{stars}$ can be straightforwardly estimated from observations \citep[e.g.,][]{meidt2012}, so that estimating $\nu$ requires only an estimate of the stellar scale height. This can be constrained either by measuring the stellar velocity dispersion \citep[e.g.,][]{Martinsson2013} or by analogy to observations of edge on disks \citep[e.g.,][]{Kregel2002,bershady,comeron2014}.

\subsubsection{Relative frequency of vertical and radial epicycles}\label{sec:nonviso}

The relative values of the vertical and radial epicyclic frequencies are determined by the galaxy potential and its variation on cloud scales.

In the case of a Mestel disk \citep{Mestel}, for example, the rotational velocity $V_{circ}$=$\sqrt{2\pi G\Sigma_{tot}R_{gal}}$ is constant given some total stellar plus dark matter surface density, $\Sigma_{tot}(r)$. Then we can write
\begin{eqnarray}
\kappa^2=2\Omega^2&=&2(V_{circ}/R_{gal})^2\nonumber \\
&=&4\pi G\Sigma_{tot}R_{gal}
\end{eqnarray}
Combining this with eq. (\ref{eq:vertfreq})\footnote{Note that eq. (\ref{eq:vertfreq}) will overestimate $\nu$ in the presence of a central bulge or bar structure, where the vertical potential gradient is weaker than in the case of the assumed disk model.  But at such a small radius $R_{gal}$$<$$2z_0$, eq. (\ref{eq:nu2}) suggest that epicyclic motions in the plane are even larger than those in the vertical direction.}, we estimate that
\begin{equation}
\frac{\nu^2}{\kappa^2}=\frac{R_{gal}}{2 z_0}\frac{\Sigma_{stars}}{\Sigma_{tot}} \label{eq:nu2}
\end{equation}

For the inner, molecule-rich parts of disks we assume that stars outmass the dark matter in the disk. We therefore adopt the approximation 
\begin{equation}
\nu^2\approx\kappa^2 R_{gal}(2 z_0)^{-1}. \label{eq:nuapprox}
\end{equation}  

Because galaxy disks have larger radial scale lengths than scale heights, from eq. \ref{eq:nuapprox} we expect that $\nu $$>$$ \kappa$ over most of the galaxy $R_{gal} $$>>$$ 2 z_0$.  For a spherical cloud of size $R_c$, the force applied by the background potential in the vertical direction $\nu^2 R_c$ will exceed the forcing $\kappa^2 R_c$ experienced in the plane.  This justifies our choice of separable potential in eq. (\ref{eq:galpot}) and our treatment of separate (non-isotropic) motions in three dimensions.  

On scales much smaller than the scale height of the background disk, however, forcing may be expected to become more isotropic and $\nu$$\sim$$\kappa$.  Here the cloud material no longer probes the disk structure and instead the background density distribution behaves more like a uniform density sphere. (The density distribution in any direction is approximately constant to zeroth order, across small distances.)  
As a result, the gradient in the potential becomes roughly the same in all directions.  

In the case of the exponential vertical distribution considered above, the variation in density across a region roughly 1/20 of the stellar scale height $z_0$ is less than 5\%.  Given values 0.25$<$$z_0$$<$0.5 kpc observed in edge-on disk galaxies (\citealt{bershady}; see Appendix B) the variation in the stellar density is negligible on scales below 10-20 pc, which is slightly less than the sizes of typical giant molecular clouds observed in nearby galaxies (i.e. \citealt{bolatto}; \citealt{hughesI}).  We therefore expect our model of 3D non-isotropic motions in $\S$ \ref{sec:netgalmotions} to be relevant on cloud-scales, although our model of 3D isotropic motions may be more applicable in the deeper interiors of clouds.  

It is worth noting that, at large enough galactocentric radius where the triaxial dark matter distribution dominates the potential, the gradient in the density and potential across the gas disk becomes so small that motions are likely to be isotropic in 3D on scales beyond the typical cloud size.  

\subsubsection{The characteristic sizes of epicycles}\label{sec:epicSizes}

We can estimate the characteristic amplitudes of the epicyclic motions in the gas by considering 
typical observed gas velocity dispersions.   
Assuming that the velocity dispersion arises only from the epicyclic motions, $\sigma^2$$\approx$$<$$\dot{x}^2$$>$, then the size of the epicycle will be:

\begin{equation}
\label{eq:rkapsig}
R_{ep}=\sigma_{gas}\kappa^{-1}
\end{equation}
This represents an upper bound on the true epicyclic radius in the presence of additional sources of motion in the gas.  

The typical velocity dispersion $\sigma_{gas}^{1kpc}$=7 km $s^{-1}$ measured in the neutral ISM on large (1 kpc) scales \citep{heiles,tamburro2009} implies epicycles as large as $R_{ep}$$\approx$ 100-200 pc, given angular velocities (and epicyclic frequencies) characteristic of star-forming disk galaxies. In reality, the velocity dispersion in the {\sc HI} also likely includes thermal contributions, implying that the dispersion due to the epicyclic motion will be somewhat smaller, lowering $R_{ep}$.  

Here we are interested in cloud-scale motions in a cold molecular medium. Adopting the lower velocity dispersion of 1-2 km/s typical of observed moderate-mass Milky Way clouds \citep[][]{heyer,mv17}, then implied epicyclic radii will instead be $R_{ep}$$\approx$20-50 pc at maximum at galactocentric radii inside one disk scalelength.  This may also be representative for epicycles in regions of larger line widths observed on cloud scales in many extragalactic systems (e.g., \citealt{hughesII}, \citealt{leroy2016}, J. Sun et al. in preparation), allowing that only a fraction of the line width may arise with epicyclic motions. 

The expected size of epicycles in the molecular gas is much closer to the scales of clouds themselves.  This suggests that epicyclic motions are not exclusively a large-scale dynamical phenomenon and remain relevant for describing motions on the cloud-scale and below.  These calculations reversely imply that the background galaxy potential induces epicyclic motions with velocities comparable in magnitude to observed velocity dispersions when applied on the scale of individual clouds.  This is considered in detail in the next section. 
Later in $\S$ \ref{sec:cloudrot} we discuss how observations can distinguish whether the observed cloud-scale motions are indeed epicyclic in nature.

\subsubsection{Net motions associated with the galaxy potential}\label{sec:netgalmotions}

We assume all cloud material has the same galactocentric orbital angular momentum so that a cloud is described by fully populated nested epicycles. Then the motion of the gas within the cloud along any given dimension will be a density-weighted integral over the full set of phases and across the set of nested epicycles in the cloud. The r.m.s. motions in the $x$, $y$, and $z$ direction will be

\begin{eqnarray}
\hspace*{-.5cm}\sigma_x^2 &=&  \int_0^{R_c} \rho (r) \left( \frac{2\pi}{\kappa}\int_0^{\frac{2\pi}{\kappa}} v_x^2 dt \right)~dr\Big/\int_0^{R_c} \rho (r) dr \label{eq:xdisp}\\
\sigma_y^2 &=& \int_0^{R_c} \rho (r) \left( \frac{2\pi}{\kappa}\int_0^{\frac{2\pi}{\kappa}} v_y^2 dt \right)~dr \Big/\int_0^{R_c} \rho (r) dr \label{eq:ydisp}\\
\sigma_z^2 &=& \int_0^{Z_c} \rho (z) \left( \frac{2\pi}{\nu}\int_0^{\frac{2\pi}{\nu}} v_z^2 dt \right)~dz \Big/\int_0^{Z_c} \rho (r) dr~. \label{eq:zdisp}
\end{eqnarray}
%
\noindent In eqs.~(\ref{eq:xdisp})-(\ref{eq:zdisp}) the time integral runs over a full epicyclic period to reflect instantaneous sampling of all parts of the motion. The integrals over $r$ and $z$ reflect the convolution of the cloud internal density distribution $\rho (r,z)$ with the nested epicyclic motions. Both the integral over the phases and the integral over the density distribution should yield pre-factors $<1$ but of order unity. We neglect them moving onward here, but will return to them in future work. Then:

\begin{eqnarray}
\sigma_x^2 &\sim& (\kappa R_c)^2 \\
\sigma_y^2 &\sim& (2 \Omega R_c)^2 \\
\sigma_z^2 &\sim& (\nu Z_c)^2
\end{eqnarray}

\noindent where we have substituted in the derivatives of eqs. (\ref{eq:xmotion}), (\ref{eq:ymotion}), and (\ref{eq:zmotion}) for $v_x$, $v_y$, and $v_z$.  Here $R_c$ is the cloud radius in the plane and $Z_c$ is the height of the cloud, which we allow to differ from $R_c$. 

We note that in the less restrictive case that clouds are composed of gas populating epicycles from around unique neighboring guiding centers, the rms velocities across the cloud in 3D are approximately the same.   In this case, the above integrals yield $\sigma_x^2$$\sim$$(\kappa R_c)^2$, $\sigma_y^2$$\sim$$(2BR_c)^2$ and $\sigma_z^2$$\sim$$(\nu R_c)^2$ (see Binney \& Tremaine 1987) assuming a uniform cloud density and adopting the cloud scale as the typical epicyclic amplitude (see $\S$ \ref{sec:epicSizes}). 

Combining the two-dimensional epicyclic motions in the galactic plane with those in the vertical direction (described in $\S$ \ref{sec:epicTraj}) we can write 
\begin{equation}
3\sigma_{gal}^2=(\kappa R_c)^2+(2\Omega R_c)^2+(\nu Z_c)^2
\end{equation}
where $\sigma_{gal}$ is the effective one-dimensional line-of-sight velocity dispersion.  
For a flat rotation curve, $\kappa=\sqrt{2}\Omega$, so that the first two terms combine to approximately yield $2(\kappa R_c)^2$ whereby

\begin{equation}
3\sigma_{gal}^2=2~(\kappa R_c)^2+(\nu Z_c)^2\label{eq:siggalnet}~.
\end{equation}

{\noindent\underline{Isotropic and non-isotropic cases}}\\

As discussed above, we expect $\nu$$>$$\kappa$ except in the centers of galaxies (or on small scales below $\approx$10-20 pc; $\S$ \ref{sec:nonviso}). In the case of the Mestel disk, for example, $\nu$ exceeds $\kappa$ at all radii $R_{gal}$$>$$2z_0$. Thus for spherical clouds with $Z_c = R_c$, we expect the last term in eq. (\ref{eq:siggalnet}) to dominate.

The relative strengths of in-plane and vertical motions also depend on the values of $Z_c$ and $R_c$. At locations where $\nu$$>$$\kappa$, flattened clouds with $R_c > Z_c$ will have vertical motions $\nu Z_c$ comparable to the motions in the plane $\kappa R_c$.  In the case where $Z_c/R_c$=$\kappa/\nu$ the motions become isotropic in all three dimensions. Outside galaxy centers, this occurs for clouds with vertical extents smaller than their in-plane size.

Moving forward, we consider the following two cases: 
\begin{equation}
\begin{cases}
3\sigma_{gal, iso}^2=3(\kappa R_c)^2 \hspace*{1.5cm}\footnotesize{\text{isotropic motions, small flattened}}\\
3\sigma_{gal}^2=2(\kappa R_c)^2+(\nu R_c)^2 \hspace*{.5cm}\footnotesize{\text{non-isotropic motions, spherical}}
\end{cases}
\label{eq:siggalcases}
\end{equation}

\noindent In the first case, isotropic internal motions require a flattened cloud geometry to balance $\nu Z_c = \kappa R_c$. The second case corresponds to spherical clouds, in which the vertical motions of the cloud will generally have large velocities than those in the plane because $\nu$$>$$\kappa$.

Note that motions in the second case exceed those in the first case. Eliminating $\nu$ in favor of $\kappa$ via eq. (\ref{eq:nuapprox}), the non-isotropic case becomes

\begin{equation}
\sigma_{gal}^2=\sigma_{gal, iso}^2+\frac{1}{3}(\kappa R_c)^2\left(\frac{R_{gal}}{2z_0}-1\right)\label{eq:casediff}
\end{equation}

\noindent which is a higher energy configuration than the isotropic case.

Note also that in the first case of flattened clouds the isotropic motions could be written $3\sigma_{gal}^2=3(\nu Z_c)^2$.  However, we prefer to invoke the cloud size in the plane, which tends to be more directly observable than the height $Z_c$ in the typically targeted galaxies we wish to compare to (although there may be instances when the vertical extent of the gas is better observationally constrained).  This is slightly different from the case in which $3\sigma_{gal}^2=3(\nu R_c)^2$, which corresponds to the standard description for isotropic gas motions in the limit of no self-gravity (see eq. \ref{eq:vertfreq1}).    We omit this case from our discussion below. It constitutes a higher energy case than either of the two cases above, with kinetic energy exceeding $\Phi_{gal}$ across the cloud. This makes the case less likely than the other two scenarios without considering an additional source of energy (e.g., feedback from star formation).  

A final note concerns the scenario in which $\nu$$\approx$$\kappa$ that we expect on scales much smaller than the characteristic variation in the background density distribution, i.e. below the vertical disk scale height $z_0$.  
This scenario would be likely on the small scales within the deep interiors of clouds.  It might also occur when the entire cloud is overall small relative to $z_0$ as a result of, e.g., a physical limit to cloud size (such as in galaxy centers M51, \citealt{colombo2014a}; MW, \citealt{oka}), or a change in the structure of the host galaxy (e.g. in galaxy outskirts, where the triaxial dark matter distribution dominates over the stellar disk).   In this limit, the non-isotropic description across a spherical region or cloud approaches the isotropic case, whereas non-spherical clouds or regions will exhibit non-isotropic motions.   

The primary two cases given in eq.(\ref{eq:siggalcases}) represent internal cloud motions associated with the galaxy potential that project into the line-of-sight differently.  In the isotropic case the line-of-sight projection of the motions $\sigma_{gal,los}$ is equal to $\sigma_{gal}$. In the second case, the projections of the vertical and in-plane motions depends on the inclination angle of the disk.  We write the general form of the line-of-sight motions due to the galaxy potential as
\begin{equation}
\sigma_{gal, los}^2=(\kappa R_c\cos{\theta}\sin{i})^2+(2\Omega R_c\sin{\theta}\sin{i})^2+(\nu Z_c\cos{i})^2
\end{equation}
where $\theta$ is the azimuthal position with respect to the disk kinematic major axis and $i$ is the inclination of the disk with respect to the line-of-sight.  

In the case of the spherical cloud with anisotropic motions, we approximate this as  
\begin{equation}
\sigma_{gal, los}^2\approx(\kappa R_c\sin{i})^2+(\nu R_c\cos{i})^2
\end{equation}
which implies that the magnitude of the recovered velocities along the line-of-sight depends strongly on galaxy inclination, especially since $\nu R_c$$>$$\kappa R_c$ on the scales of typical clouds.

\subsubsection{Departures from axisymmetry in the galaxy potential}\label{sec:spiralkappa}

So far we have exclusively considered the case of an axisymmetric galaxy potential.  But dynamical features such as bars and spiral arms present in the stellar density distribution locally contribute an additional gradient in the gravitational potential, which alters the gas kinematics in the form of the epicyclic motions described in $\S$ \ref{sec:epicycles}.  

To estimate the local change in the epicyclic frequency in the presence of a perturbation to the galaxy gravitational potential, we describe the perturbation following \cite{BT}. In the case of a bar rotating with angular pattern speed $\Omega_p$ we write the perturbed potential as $\Phi_1(\theta,t)$=$\Phi_b e^{im(\Omega-\Omega_p)t}$ where $\Phi_b$ is independent of radius,  as in the case of a weak bar (e.g. \citealt{sell10}).  Likewise, we describe spirals as wave solutions to the perturbed equations of motion in the tight-winding limit (WKB approximation) that propagate with wavenumber $k$ and pattern speed $\Omega_p$.  For a density perturbation $\Sigma_1=\Sigma_a e^{ikR-\Omega_pt}$ the corresponding density wave potential perturbation has the form $\Phi_1(r,\theta,t)$=$\Phi_a e^{ikR-\Omega_pt}$, where $\Phi_a=2\pi \Sigma_a/k$ via Poission's equation.  

In both cases, the perturbed epicyclic frequency is the circulation frequency around the pattern, i.e. $m(\Omega-\Omega_p)$.  As the total galaxy gravitational potential is the sum of the unperturbed and perturbed components, $\Phi$=$\Phi_0(r)+\Phi_1(r,\theta,t)$,  the epicyclic frequency becomes
\begin{equation}
K^2=\kappa^2+m^2(\Omega-\Omega_p)^2\label{eq:kappaspiral}
\end{equation}
where $\kappa$ is the epicyclic frequency in the unperturbed, axisymmetric potential.  Here we label epicyclic motions in the presence of non-axisymmetric potentials $K$ but elsewhere we refer to epicyclic motions generally as $\kappa$ unless otherwise noted.  

Far enough from the corotation radius (where $\Omega$=$\Omega_p$) of a bar or a two-armed spiral (with $m$=2) 
\begin{equation}
K^2\approx\kappa^2+4\Omega^2.  \label{eq:kappaspiralEst}
\end{equation}
Generally, for small displacements $\epsilon_r$ from the corotation radius $R_{CR}$ of the pattern
\begin{equation}
K^2\approx\kappa^2+4\Omega^2\left(\frac{\epsilon_r}{R_{CR}}\right)^2.  \label{eq:spiralepicApprox1}
\end{equation}

\subsection{Internal motions including gas self-gravity}\label{sec:selfgrav}

Gas motions should also reflect the potential defined by the cloud material. For a cloud of mass $M$ with a volume density profile $\rho\propto r^{-k}$ we write the potential as
\begin{eqnarray}
\Phi_{c} = \frac{3}{5}a_k\frac{G M_c}{R_c} \label{eq:cloudpot}
\end{eqnarray}
following \cite{bertoldimckee} where
\begin{equation}
a_k=\frac{(1-k/3)}{(1-2k/5)}.
\end{equation}  
In the case of a homogeneous (uniform density) cloud, $a_k$=1.  For an isothermal cloud with $\rho\propto R_c^{-2}$, $a_k$=5/3. 

\subsubsection{Motions due to self-gravity}\label{sec:equiselfgrav} 

For an isolated cloud in energy equipartition, the potential defined by the weight of the cloud is matched by the kinetic energy.  
We assume that the kinetic energy released by self-gravity is in
the form of infall motions that we describe with the one-dimensional
velocity dispersion
\begin{equation}
\sigma_{sg}=\sqrt{2 (\pi a_k/5) GR_c\Sigma_c}.\label{eq:sigSG}
\end{equation}
(where $R_c$ is the cloud `radius' and $\Sigma_c$ is the cloud surface density).  

Here $\sigma_{sg}$ is distinct from the observed gas velocity dispersion $\sigma_v$.  $\sigma_{sg}$ specifically refers to the motions that 
 arise from the gas self-gravity. Following \cite{vs08}, \cite{ballesteros} and \cite{ibanez} we envision these motions as the response of the gas to collapse under its own weight.  But we also allow that such motions may be only part of the full spectrum, as additional, turbulent motions can arise through a variety of mechanisms, including feedback or instabilities on larger scales.  

In an isolated, self-gravitating cloud with no angular momentum we expect the motions to be quasi-isotropic. Simulations of cloud-scale hierarchical gravitational collapse tend to find multiple internal centers of collapse, yielding chaotic but quasi-isotropic velocity fields (\citealt{ballesteros}).  Likewise, in the `gravitational cascade' picture of \cite{field08}, the kinetic energy released during collapse is in the form of quasi-isotropic motions. Because the motions due to self-gravity are not expected to be ordered, any net rotation should stem from the epicyclic motion predicted above.  The impact of non-zero angular momentum and the properties of turbulence on the net internal kinematics will be discussed later in $\S$ \ref{sec:cloudrot}. 

\subsubsection{Net gas motions}\label{sec:netmotions}

We envision two scenarios for gas motions framed by the combination of gas self-gravity and the background galaxy potential.  In the first case, we hypothesize that the collapse response of gas to its own self-gravity will release the kinetic energy associated with eq. (\ref{eq:sigSG}), even in the presence of motions induced by the galaxy potential. In this case, we expect all clouds to develop internal motions of magnitude $\sigma_{sg}$, which then add to the coherent motions already due to the galaxy potential.   

We write the net motions associated with both cloud collapse and the background galaxy potential as the sum in quadrature of $\sigma_{sg}$ and $\sigma_{\rm gal}$,

\begin{equation}
\sigma^2=\sigma_{sg}^2 + \sigma_{gal}^2 = \sigma_{sg}^2+\kappa^2 R_c^2 \label{eq:vdisp}
\end{equation}
where the latter expression adopts the isotropic case for motions due to the galaxy potential.

This assumption, which should eventually be tested numerically, implicitly assumes that motions associated with gas self-gravity and with the background galaxy potential are uncorrelated, i.e. the gravitational potential of the gas cloud is assumed to be distinctly different (statistically independent) from the local gravitational potential defined by the background galaxy disk.  Thus, even in the case that infall motions at a given location within a cloud are balanced by epicyclic motions, the overall motions in the cloud associated with either gravitational potential add in quadrature.    

In the limit that the background galaxy exerts negligible influence, the expression in eq. (\ref{eq:vdisp}) reduces to
\begin{equation}
\sigma^2=\sigma_{sg}^2
\end{equation}
with internal cloud motions reflecting only the self-gravity of the gas.  As we discuss below, this scenario should be distinguishable from the case where the background potential becomes significant, so that $\sigma_v$ is not equal to $\sigma_{sg}$ (see eq. [\ref{eq:vdisp}]).  
The latter should lead to observed virial parameters largely in
excess of the $\alpha_{obs}$=2 expected for marginally bound or
free-falling self-gravitating gas.  More, it should lead to an observable dependence of the observed virial parameter on the background potential.

Eq. (\ref{eq:vdisp}) represents a hypothesis, which we refer to as {\em energy equipartition}. We assume that motions associated with gas self-gravity emerge independent of the background galaxy potential. Even in the case that epicyclic motions play a role in collecting the gas or help support the cloud, we assume that motions of $\sigma_{sg}$ will still develop. Energetically, we argue that 
cloud motions should reflect both the potential of the galaxy and potential of the cloud.

A conceptual alternative is that gas motions reflect {\em self-regulation} to some equilibrium dynamical state. Self-regulation by star formation has been invoked many times, e.g., \cite{ostriker} and \cite{ostrikerShetty}.  In this second case, motions arising with feedback from star formation are thought to provide a time-averaged balance against the cloud's self-gravity and keep the cloud near a stable dynamical state.\footnote{
Numerical simulations performed at the cloud scale often show that either the effect of feedback is to 
disrupt or evaporate clouds or it is incapable of preventing cloud collapse, rather than acting as a way to maintain clouds in near-equilibrium (e.g., \citealt{dale2012}; \citealt{colin2013}).  
}  
Thus we might expect the {\em net} cloud motions to approximately equal $\sigma_{sg}$.  As energy dissipates, the cloud becomes self-gravitating, leading to further star formation and feedback that ultimately generates more more motions.  

In the self-regulation scenario, the motions due to the galactic potential may still be present and our calculations above remain relevant. These may contribute support to the cloud, lowering the amount of feedback necessary to keep the gas near a fixed dynamical state. As we discuss below, in the radial direction, this resembles the case of Toomre $Q$ self-regulation.

In the next section we discuss observational tests that could distinguish these hypotheses.

\subsection{The relative importance of the cloud potential and the local background galaxy potential}\label{sec:forcebalance}

It will often be of interest to compare the self-gravity of a cloud to the gravitational potential defined by the host galaxy. We use the motions that arise due to each potential as a proxy for the potential strength. We estimate the ratio of forces due to the galaxy potential to those due to self-gravity as
\begin{eqnarray}
\frac{1}{\gamma^2}&=&\frac{\nabla\Phi_{gal}}{\nabla\Phi_{c}}\nonumber \\ 
&=&\frac{3\sigma_{gal}^2}{3\sigma_{sg}^2}\nonumber \\ 
&\approx&\frac{3\kappa^2 R_c}{2 (3 a_k/5)\pi G \Sigma_c}\nonumber \\
&\approx&\left(\frac{\kappa R_c }{\sigma_{sg}}\right)^2 \label{eq:balance}
\end{eqnarray}
where we have adopted the isotropic case in the final two lines. Here $\kappa$ would be replaced by $K$ in eq. (\ref{eq:kappaspiral}) in the presence of, e.g. a spiral potential perturbation.    

Motions due to the galaxy potential balance the potential energy of the cloud when $\gamma$=1.  If motions due to the collapse of the cloud are also present, then gas can be said to be self-gravitating only when $\sigma_{sg}$$>$$\kappa R_c$ or $\gamma$$>$1. In this case, motions due to the potential of the cloud exceed those due to the galaxy potential.

The value of $\gamma$ for which the gas is strongly self-gravitating is presumably nearer to $10$ than 2. In this case the self-gravity exceeds the galaxy potential by a factor $\approx$100. In paper II we calibrate the value of $\gamma$ associated with the onset of star formation based on observations of clouds in the Solar Neighborhood.

\subsubsection{Analogy to Toomre Q}

The balance between the local galactic potential and the self-gravity of a gas cloud resembles the considerations behind the Toomre $Q$ parameter \citep{toomre}. Formally expressed as
\begin{equation}
Q=\frac{\sigma_{v}\kappa}{\pi G\Sigma} \label{eq:toomre}~,
\end{equation}
$Q$ measures the stability of gas disks against collapse, with $Q < 1$ indicating instability. Here $\sigma_v$ refers to the velocity dispersion due to random motions in the radial direction. $Q$ is defined by matching the scale at which gravity balances internal motions with the scale at which gravity balances the Coriolis force.

The Toomre parameter can be used to approximate gas stability at cloud scales by combining eqs. (\ref{eq:sigSG}) and (\ref{eq:toomre}), 
\begin{eqnarray}
Q_{c}&=&\frac{2 \sigma_{v}\kappa R_c}{(5/a_k)\sigma_{sg}^2} \sim \frac{2 a_k}{5} \left(\frac{\sigma_v}{\sigma_{sg}}\right) \left(\frac{\kappa R_c}{\sigma_{sg}}\right)~. \label{eq:qc}
\end{eqnarray}
$Q_c$=1 corresponds to a balance between the forces experienced by clouds. From eq. (\ref{eq:qc}), we see that our condition for strongly self-gravitating clouds, $\sigma_{sg}$$>>$$\kappa R$, will typically correspond to $Q_c$$<<$ 1.  But note that, depending on $\sigma_v$ compared to $\sigma_{sg}$, clouds with $\sigma_{sg}$$>>$$\kappa R$ might still be stable against collapse, with $Q_c$$\approx$1.  These clouds would lie in the regime described by the Jean's criterion, where rotation plays a negligible role \citep[e.g.,][]{Schaye2004} and clouds are stabilized by pressure support.  

Despite the conceptual similarity to the force balance expressed in the previous section, the Toomre $Q$ parameter notably applies differently in practice. $Q$ describes the balance of forces on a particular spatial scale in the disk plane (associated with the most unstable mode in the disk), whereas our self-gravitation criterion is designed to capture gravitational forces at any scale throughout the gas, including within clouds, and considers all three spatial dimensions. Modifications to the Toomre criterion for gas in the presence of an external potential \citep{jog13} and due to finite disk thickness \citep{romeo}, better approximate the scenario our model is intended to describe.  We will use the more generic criterion implied by our model, developed expressly for describing the cloud-scale balance of forces in three dimensions.   

\subsection{Comparisons to other models }\label{sec:othertheory}

\subsubsection{The role of the epicycle }

Through the Toomre criterion, epicyclic motion plays a role in several models that describe the influence of the host galaxy on the ISM and star formation.  In many of these models, star-forming gas is thought to self-regulate so that $Q\approx$1 (e.g., \citealt{ko01}; \citealt{ko06}; \citealt{bournaud}; \citealt{krumBurk}). Our model differs from these in several important ways. First, we do not invoke self-regulation to any particular $Q$ value. Second, we consider epicyclic motions as pertaining to the material within a cloud.  Previous focus has considered clouds as ballistic objects that themselves undergo epicyclic motions.  In this way, the epicycle introduces time-dependent tidal forces \citep{das} that represent a potential source of cloud heating.  In many scenarios, galactic motions determine where clouds form through the Toomre instability, after which universal properties effectively instantaneously set in.  In our treatment, the same galactic motions envisioned by Toomre extend to within the cloud interior.  

\subsubsection{Vertical equilibrium and self-regulated star formation}

Our picture resembles the model proposed by \cite{ostriker} and \cite{ostrikerShetty}, 
in which the fraction of gravitationally bound, star-forming gas is linked to vertical dynamical equilibrium.  Equilibrium is achieved through a balance between the vertical weight of the gas due to the stars and dark matter and the (turbulent and thermal) pressure in the diffuse gas originating with the feedback from star formation. The model invokes star formation self-regulation to maintain this equilibrium. Too little star formation and the gas disk is under-pressurized, leading to collapse and an increase in star formation, which raises the pressure and subsequently reduces star formation. This helps explain the almost universal inefficiency of star formation in normal  galaxies (\citealt{ostriker}; \citealt{hopkins}).

The model proposed here differs in two ways. First, we treat motions in the plane, in addition to the vertical direction. This synthesizes the two main mechanisms proposed to regulate star formation: vertical pressure equilibrium (\citealt{ostriker} and \citealt{ostrikerShetty}) and shear and Coriolis forces as parameterized by Toomre Q$\approx$1 (e.g., \citealt{koyama}; \citealt{hunter}; \citealt{hopkins}).  But, whereas gas disks have been considered either razor thin (with no vertical structure) or composed of ballistic self-gravitating clouds, in this new model galaxy-induced motions pervade the cloud interior.   

Second, we do not invoke feedback-driven self-regulation. Instead, we envision cloud-scale motions as generated by the combination of collapse due to self-gravity and the influence of the galaxy potential. This framework therefore does not assume a particular critical state {\em a priori} but  considers the velocity dispersion in the gas as implied by the equipartition of energy between motions and the relevant potentials.

\subsubsection{The combined influence of feedback and gravity}

The formalism presented here can be easily modified to include the energetic contribution from other potential sources of motion, including mechanical feedback from star formation, which is an important barrier to instability and collapse in many models of star formation (i.e. \cite{krumholz05}; \citealt{hennebelle}; \citealt{ostriker}; \citealt{padoan}).  

Simulations show that feedback can internally support clouds in rough virial equilibrium (e.g. \citealt{goldbaum}; \cite{zamora}) and maintain disks in time-averaged vertical pressure equilibrium (i.e. \citealt{ostrikerShetty}; \citealt{shettyOstriker}). This implies that motions associated with feedback could contribute an additional energy term that roughly matches self-gravity.

How these motions should be combined with those due to collapse is not clear.  In an equilibrium picture, mechanical feedback might cancel out collapse, removing one term from our net motions and leaving another of the same magnitude.  At maximum, the motions will combine.  Given the plausible magnitude of feedback from simulations, 
we expect that $\sigma^2$ could be raised by as much as $\approx\sigma_{sg}^2$. The motions due to the galaxy potential are roughly the same as those associated with self-gravity on the cloud scale (Fig. 2), so the internal cloud velocity dispersion due to the combination of gravitational forces and feedback might appear as large as $\approx\sqrt{3}\sigma_{sg}$.  

It is worth noting that feedback from star formation may provide an important avenue for removing the ordered nature of the epicyclic motions and promoting dissipation in the gas through shocks and collisions. But feedback may also help distribute the gas to large scales, where it again becomes dominated by orbital motions. This energy might then re-emerge as cloud-scale motions as the gas collapses again. The mechanical energy from feedback may also help drive evolution in the orbital energy distribution of the gas. This, too, would ultimately create the pattern of coherent epicyclic motions modeled here.

Such links remains speculative. Simulations combining a realistic, rotating galaxy potential, feedback, and gas self-gravity will help resolve the nature (coherent, chaotic, turbulent) and magnitude of motions on different spatial and time scales.

\subsubsection{Relation to turbulence}\label{sec:turbulence}

In the limit of a dominant external galaxy potential, gas should exhibit an ordered pattern of coherent epicyclic motions.  Other sources of motion, like magnetic fields and mechanical feedback from star formation, will change this picture. 
Additional sources of energy can disrupt the coherent motions predicted under the influence of gravity, leading to shocks and local instability in the gas that can convert orbital energy in to turbulent motions, in particular.

A number of mechanisms have been proposed to transfer kinetic energy from coherent to turbulent motions. This could be possible via one, or several, of the large-scale instabilities proposed by, e.g., \cite{sbalbus}, \cite{kim03}, \cite{wada}, \cite{vs06} and \cite{kko06}. If the instability required to generate turbulence is gravitational in nature, this could offer a more pervasive mechanism to convert galaxy-induced motions into turbulence. The process of cloud formation, itself, may be one of gravitational collapse from the surrounding medium, such as envisioned by \cite{vs06} and earlier by \cite{ko01}. In this case, gravitational motions would be converted into turbulence at the cloud scale. But if only a fraction, and not the whole cloud, is collapsing, we might expect the motions introduced by the galaxy to remain coherent until gravitational instability sets in, e.g. on the smallest scales at the highest densities.  

An alternative path to turbulence could arise with the dissipative, compressible nature of the gas, which leads to the development of small-scale shocks that dissipate the kinetic energy (converted either into heat or turbulence).  We note, though, that the coherent, epicyclic nature of the motions would tend to lengthen the dissipation timescale.

Numerical simulations are the ideal setting to identify the mechanisms and scales at which turbulence is generated from gravitational motions.  This requires including an external potential, with shear (from differential rotation) and Coriolis forces, in local turbulent box models, or perhaps tuning external driving to approximate the sustained, coherent motions predicted in axisymmetric disks (presumably resembling solenoidal driving; although motions resulting from passage through a spiral arm, for example, could introduce a more compressive component).  

Larger-scale high resolution simulations that model the dynamics of the gas in the context of a host galaxy potential (e.g. \citealt{ko02}; \citealt{tasker}; \citealt{dobb06}; \citealt{dobbsPringle}; \citealt{renaud13}; \citealt{smith}) are optimal for recognizing the scales at which coherent motions develop into turbulence.  Although the state of these simulations is changing rapidly, so far few include self-gravity or achieve the resolution necessary to capture motions much below the cloud scale, which is key for identifying how and where clouds decouple from bulk motions in the rotating gas disk.  We note that cloud-scale gas motions consistent with those predicted here to originate under the influence of host galaxy are recognizable in the AMR simulations studied by Utreras et al., (in prep.).  Observationally, the transition from galaxy-induced coherent motions within clouds to turbulence and/or to collapse motions may be recognizable using dendrograms \citep{ros08} or similar diagnostics of the kinematic and spatial structure of molecular gas.  Later in $\S$ \ref{sec:cloudrot} we discuss the signatures of coherent motions on cloud scales.  

\begin{figure*}[t]
\begin{center}
\begin{tabular}{cc}
\includegraphics[width=0.45\linewidth]{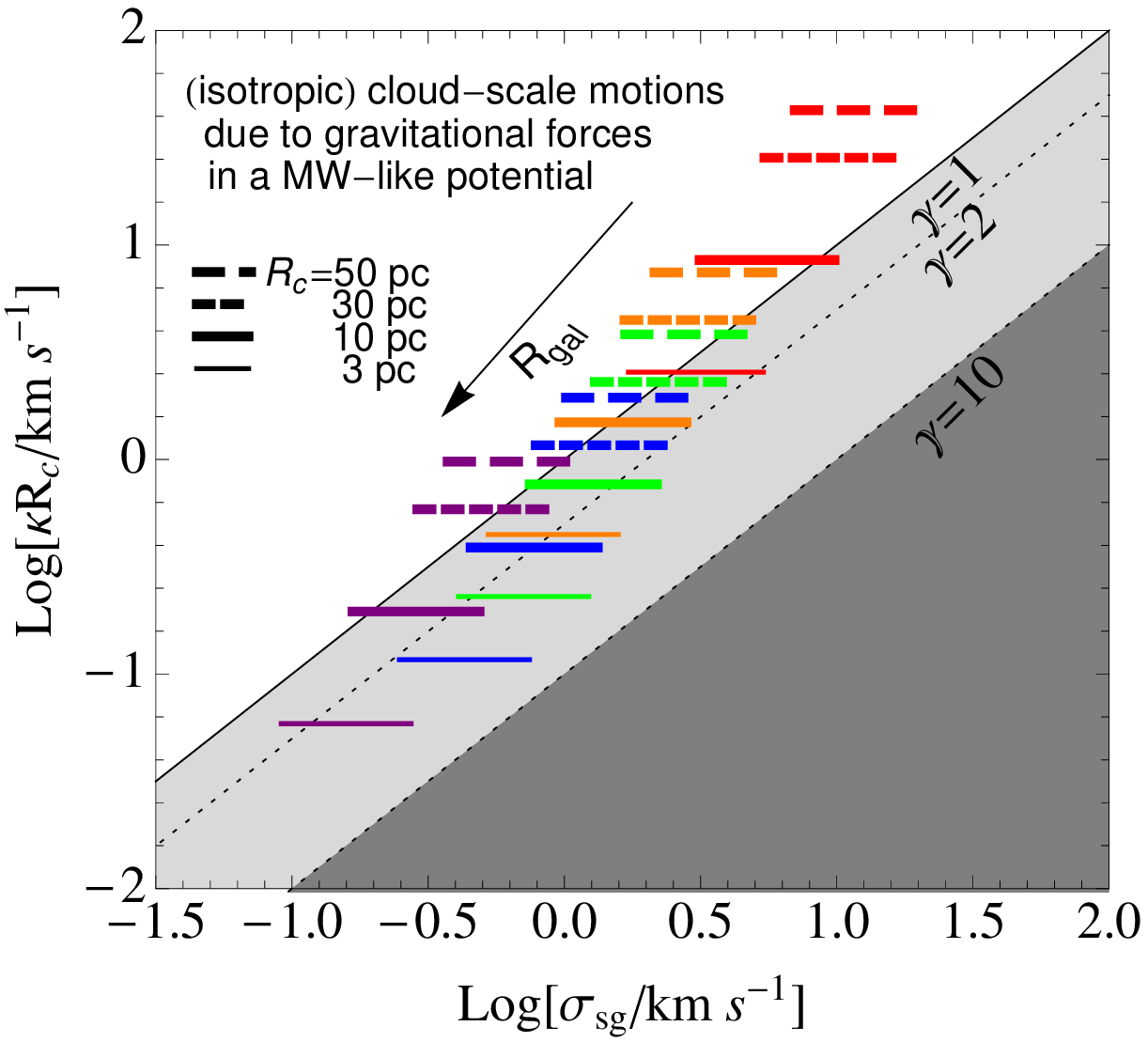}&\includegraphics[width=0.45\linewidth]{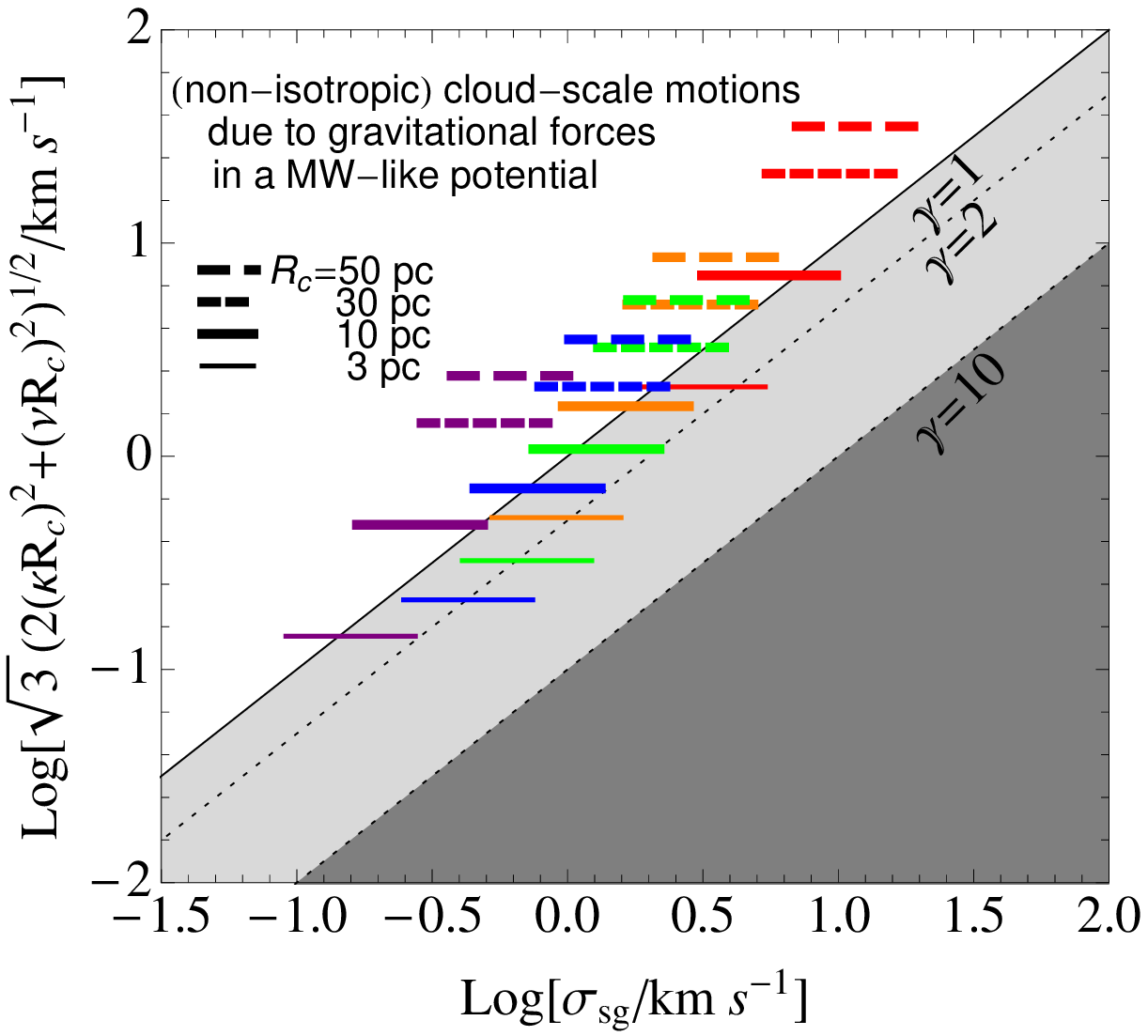}
\end{tabular}
\end{center}
\caption{Comparison of the relative strengths of gravitational forces on cloud scales in a MW-like galaxy.  The galaxy gravitational potential is represented by epicyclic motions in either the isotropic (left) or non-isotropic (right) cases on the vertical axis.  On the horizontal axis the motions $\sigma_{sg}$ due to the gas self-gravity are shown.   
In each panel, a series of five sets of lines portray values at increasingly large galactocentric radius $R_{gal}$=0.015, 0.5, 1, 2 and 4 $R_e$ from red to purple, right to left.  At fixed $R_{gal}$, the width of the line denotes a factor of ten spread in gas surface density around the average value suggested by the model described in Appendix \ref{sec:appendix}.  The three different line stylings show values for four different cloud radii $R_c$=3, 10, 30 and 50 pc.  
In the shaded region at the right of each panel, three diagonal lines highlight regimes of different relative force strength $\Gamma_{sg}$=$\gamma_{sg}^{-1}$$\approx$$\kappa R_c/\sigma$$\approx$$Q_c$.  When $\Gamma_{sg}$=1, gravitational forces balance.  When $\Gamma_{sg}$$<<$1, the gas is self-gravitating.   \vspace*{0.35cm} 
 }
\label{fig:paramspaceMW}
\end{figure*}

\section{Quantitative predictions of the model: net gas motions}\label{sec:quant}

The magnitude of the velocities due to the galaxy potential is set by the epicyclic frequencies $\kappa$ and $\nu$.  Because $\kappa$ and $\nu$ vary throughout galaxies (and from galaxy to galaxy), we expect the importance of the galaxy potential to vary with respect to gas self-gravity. To examine how the balance of gravitational forces changes throughout realistic molecular cloud populations, we compare a model for the expected behavior of the motions due to the galaxy $\sigma_{gal}$ in eq. (\ref{eq:siggalcases}) with those due to self-gravity.

In this illustration, the dependence of $\sigma_{gal}$ on location in a galaxy is determined by the galaxy's stellar mass.  The galaxy mass is empirically linked to the shape and maximum of the galaxy's rotation curve via global scaling relations. Given a total stellar mass, the model yields $\kappa$ as a function of radius according to eq. (\ref{eq:kappafromomega}).  We also estimate the frequency of vertical motions $\nu$ in relation to $\kappa$ using eq. (\ref{eq:nuapprox}) and adopting a stellar scale height appropriate for the mass of the given galaxy (see Appendix B). Along with the cloud scale, $R_c$, this yields an estimate for the magnitude of the net epicyclic velocities.

The gas self-gravity also varies. For a given $R_c$, the motions due to self-gravity $\sigma_{sg}$ are set by the assumed cloud surface density. To account for changing cloud populations, we allow the gas surface density on cloud scales to decrease with galactocentric radius, again following an empirically-motivated model (Appendix \ref{sec:appendixGas}). This model is motivated by observations of the MW \citep{mv17} and can span several decades in cloud surface density. For most of our calculations, the outermost galactocentric radius is intentionally located near the edge of typical molecular disks in normal star-forming galaxies (e.g., \citealt{schruba}; see Appendix \ref{sec:appendixGas}).  

\subsection{Motions in a Milky Way-like potential}\label{sec:quantplots}

Figure \ref{fig:paramspaceMW} shows predictions for a galaxy with a stellar mass equal to that of the Milky Way (MW). We plot motions associated with the galaxy potential as a function of the dispersion $\sigma_{sg}$ associated with self-gravity. To estimate $\sigma_{gal}$, we use the empirically-based rotation curve described above and the mass of the MW. $\sigma_{sg}$ also varies, in this case according to the fully modeled range of cloud surface densities $\Sigma_c$=1-2000 $M_\odot pc^{-2}$ with $\Sigma_c$ decreasing across the radial range $R_{gal} = 0.015{-}4~R_e$. At each $R_{gal}$ we show calculation for four cloud sizes, $R_c$=3, 10, 30, 50 pc.   

Figure \ref{fig:paramspaceMW} shows that motions due to the galaxy potential match or exceed those expected for self-gravity across a wide range of galactocentric radii, cloud surface densities, and cloud sizes. In the non-isotropic case (right) motions due to the external potential become even stronger at large galactocentric radii, reflecting strong vertical motions. Overall, the calculation argues strongly that forces due to the galactic potential, including Coriolis forces, are not negligible at cloud scales.

This assessment holds at other galaxy masses, as illustrated in Figure \ref{fig:paramspace} in  Appendix \ref{sec:appendixnu}.  The MW is on the high end of the mass range spanned by the local galaxy population. We expect the galaxy gravitational potential to become even more important relative to self-gravity in less massive disks. This effect will be compounded by the fact that lower mass tend to host clouds that are less massive and of lower surface density than clouds in more massive galaxies \citep{ros05}, reducing the strength of the gas self-gravity on the cloud scale.

Note that the importance of the galaxy potential does depend on the assumed cloud properties. For example, clouds situated the innermost $R_{gal}$ with surface densities $\Sigma_c$$\lesssim$200 $M_\odot pc^{-2}$, lower than considered in Fig.  \ref{fig:paramspaceMW}, will have $\kappa R_c$$>>$$\sigma_{sg}$. Meanwhile in the outer disk, clouds with surface densities $\Sigma_c$$\gtrsim$10 $M_\odot pc^{-2}$, also not shown, can be strongly self-gravitating, with $\kappa R_c$$<<$$\sigma_{sg}$. Systematic variations in cloud size with galactocentric radius, such as the decrease from the disk to the center of the MW (e.g. \citealt{henshaw}; and see Meidt et al., in prep.) will also affect the relative importance of the galaxy potential.

The presence of stellar dynamical features like spiral arms also enhance epicyclic motions relative to $\sigma_{sg}$.  As modeled in $\S$ \ref{sec:spiralkappa} and illustrated in Figure \ref{fig:paramspace} (Appendix \ref{sec:appendixnu}), spiral arm perturbations introduce locally large gradients in the gravitational potential, raising the epicyclic frequency (eq. \ref{eq:kappaspiral}).  In these environments, the background host galaxy makes an increased contribution to observed motions, which are thus raised above the level expected due to only gas self-gravity.  

\cite{BP09II} also emphasize the importance of the external potential defined by the background disk galaxy compared to the self-gravity of clouds, with a treatment that includes both tidal and Coriolis forces, thus making it most directly analogous to the framework we develop here.  As highlighted in Figure \ref{fig:paramspace}, our empirically-motivated description of the galaxy potential suggests that it can be similarly strong or stronger than gas self-gravity as compared to the model parameters chosen by \cite{BP09II}, particularly towards galaxy centers, which were omitted in that study.  

\subsection{Characterizing the dynamical state of the gas on cloud scales}\label{sec:dynstate}

We expect the background galaxy potential to play an important role in internal cloud motions. Here we describe how motions induced by the galaxy should be manifest in several observational diagnostics of cloud dynamical state.  In $\S$ \ref{sec:datacomp} we show that the trends predicted by the model agree with observations.

\subsubsection{The linewidth-size relation coefficient}\label{sec:diag1}

One of the contemporary diagnostics of the dynamical state of molecular clouds is the dependence of the 'size-linewidth relation coefficient' 
\begin{equation}
a=\frac{\sigma_{obs}^2}{R_c}
\end{equation}
on cloud surface density $\Sigma_{c}$.  This diagnostic is used in the literature to highlight the sensitivity of clouds to their environment (e.g. \citealt{heyer}; \citealt{field}; \citealt{ballesteros}; \citealt{leroy2015}; \citealt{hughesIII}).  

\begin{figure}[t]
\begin{tabular}{c}
\includegraphics[width=0.965\linewidth]{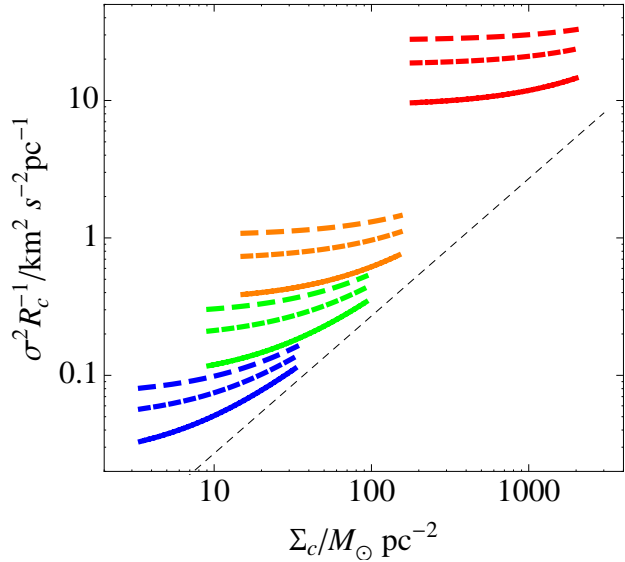}
\end{tabular}
\caption{Predicted trends in the `size-linewidth relation coefficient' $a=\sigma^2R_c^{-1}$ vs. cloud surface density $\Sigma_{c}$, where $\sigma$ reflects motions due to both gas self-gravity and the background galaxy potential.  (Other sources of motion, such as feedback and magnetic fields, have been omitted here for illustration.) The stellar mass of the galaxy in this example is $M_\star$=$10^{10.5} M_\odot$.  
Five sets of three curves are shown, each at a different galactocentric radius: $R_{gal}$=0.015$R_e$ (red), $R_{gal}$=0.5$R_e$ (orange), $R_{gal}$=1$R_e$ (green) and $R_{gal}$=2$R_e$ (blue).   
At each $R_{gal}$ trends for three cloud sizes are shown: $R_c$=15 pc (solid), $R_c$=30 pc (short dash) and $R_c$=45 pc (long dash).  The thin diagonal dashed line shows the trend predicted in the case of only self-gravity.  
Large clouds are offset to higher $a$ at fixed $\Sigma_{c}$ when gas self-gravity is relatively weak compared to the galaxy potential (i.e. at low $\Sigma_{c}$ and/or large $R_{gal}$).   \vspace*{0.35cm}   
}
\label{fig:cloudReln}
\end{figure}

Figure \ref{fig:cloudReln} shows an idealized sketch of $a$ vs. $\Sigma_c$ for a model cloud population spanning a range of galactocentric radii. Here, $\sigma_{obs}^2 = \sigma_{sg}^2+ \sigma_{gal}^2$, the quadrature sum of both types of motion in eq. (\ref{eq:vdisp}). For this figure we adopt a galaxy mass slightly lower than the MW in order to highlight the stronger effect of the potential in these targets. See Appendix \ref{sec:quantplots2} for more detail. Similar to Figure \ref{fig:paramspaceMW}, the model cloud surface densities decrease with galactocentric radius (see Appendix \ref{sec:appendixGas}), but with a large range of surface densities still present at each radius.

In a plot like Figure \ref{fig:cloudReln}, isolated, self-gravitating clouds follow a relation like the black dotted line. Incorporating motions due to the galaxy potential, our model predicts a strong deviation from this trend. Clouds show higher line width at a given surface density than for the pure self-gravitating case, with low $\Sigma_c$ clouds most affected.

At a given $R_{gal}$, clouds with low $\Sigma_c$ and high $R_c$ exhibit the largest deviations from the self-gravitating case. The size dependence  may seem surprising given that $a\propto R_c^{-1}$.  However, the dependence arises when $\sigma$ is dominated by $\sigma_{gal}$, which increases with $R_c$.

The calculations that produce Fig. \ref{fig:cloudReln} show strong variation of $a$ with galactocentric radius. We expect $a$ to be preferentially elevated at small galactocentric radius, where the galaxy potential dominates gas self-gravity.  Clouds in the MW exhibit a similar trend \citep{mv17}.  

\subsubsection{The virial parameter}\label{sec:virial}

The virial parameter $\alpha$ assesses the ratio of the kinetic and potential energies in a cloud. Models of the turbulent ISM link $\alpha$ to the onset of star formation (e.g., \citealt{padoanRev}). $\alpha$=1 indicates that cloud is in virial balance. $\alpha$=2 represents energy equipartition. 

Typically, the virial parameter is constructed from the observed velocity dispersion, a quantity we refer to as $\alpha_{obs}$. The velocity dispersion used to calculate $\alpha_{obs}$ will include motions due to the galaxy potential. To consider the balance only between the random, internal motions and the cloud potential, which is likely still of interest, we write

\begin{eqnarray}
\alpha_{corr}&=&\left(3\sigma_{obs}^2-3\kappa^2 R_c^2\right)\left(\frac{3 G M}{5 R_c}\right)^{-1}\nonumber\\
&=&\alpha_{obs}-\frac{3\kappa^2 R_c^2}{3 G M/(5 R_c)}\nonumber\\
&=&\alpha_{obs}-\left(\frac{\kappa R_c}{\sigma_{sg}}\right)^2,  \label{eq:alphagal}
\end{eqnarray}
for the isotropic case and 
\begin{eqnarray}
\alpha_{corr}&=&\left(3\sigma_{obs}^2-2\kappa^2 R_c^2-\nu^2 R_c^2\right)\left(\frac{3 G M}{5 R_c}\right)^{-1}\nonumber\\
&\approx&\alpha_{obs}-\left(\frac{\nu R_c}{\sqrt{3}\sigma_{sg}}\right)^2  \label{eq:alphagal2}
\end{eqnarray} 
for spherical clouds in the anisotropic case. The resulting $\alpha_{corr}$ can then be used to compare the cloud self-gravity to its random internal motions. Without this correction, the observed virial ratio $\alpha_{obs}$ will overestimate the ratio $\alpha_{corr}$ by at least $\approx\kappa^2 R_c^2/\sigma_{sg}^2$.

\begin{figure}[t]
\includegraphics[width=0.965\linewidth]{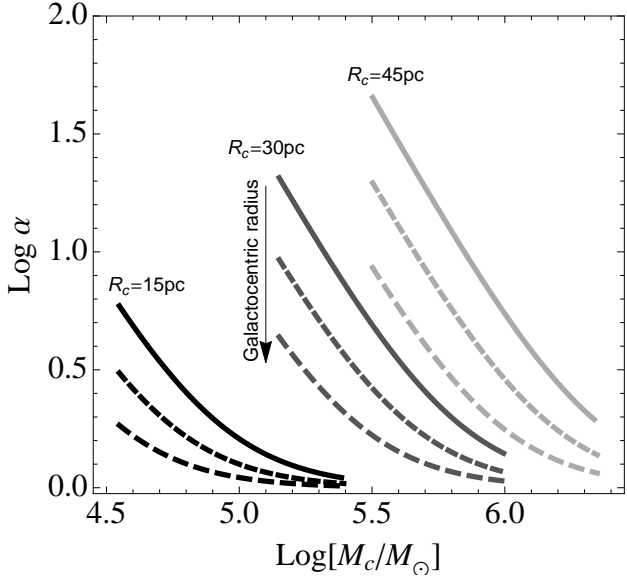}
\caption{Example of the behavior of the virial parameter $\alpha$ constructed from the observed velocity dispersion in the case that the cloud's true virial balance includes self-gravity and the influence of the local galaxy potential so that $\alpha_{gal}$=1 in eq. (\ref{eq:alphagal}).  (Other sources of motion, such as feedback and magnetic fields, have been omitted here for illustration.) Each curve traces out the variation expected for cloud surface density increasing from 50 to 350 M$_\odot$ pc$^{-2}$ (left to right). Measurements at three cloud scales (from left to right; dark to light gray: 15, 30 and 45 pc) are shown.  Three curves at fixed scale show the trends at three different galactocentric radii (from solid to dashed: 0.5 $R_e$ to 1.0 $R_e$).  In this sketch, the scale length $R_e$ and the epicyclic frequency $\kappa$ associated with a galaxy of stellar mass $M_{\star}$=10$^{10.5}$ M$_\odot$ (according to our empirically-based rotation curve model) is adopted.  
\vspace*{0.35cm}  }
\label{fig:alpha}
\end{figure}

Figure \ref{fig:alpha} illustrates how the virial ratio $\alpha_{obs}$ would appear for isotropic clouds in true virial equilibrium with $\alpha_{corr}$=1 in eq. (\ref{eq:alphagal}).  Low mass clouds with low surface densities appear systematically more stable against gravitational collapse. In this example, the apparent discrepancy is larger at smaller galactocentric radii, where $\kappa R_c$ becomes stronger. 

The strong radial dependence of $\kappa$ implies that $\alpha_{obs}$ should show signs of variation with radius.  We note that an equally likely recognizable trend (not illustrated) is azimuthal variation reflecting an increase within bars and in spiral arms compared to inter-arms where $\kappa$ increases to $K$ as explored in $\S$ \ref{sec:spiralkappa}.  

The difference between $\alpha_{corr}$ and $\alpha_{obs}$ means that gas may be in a true state of virial equilibrium even when $\alpha_{obs} > 1$. It remains unclear, however, whether $\alpha_{corr}$ or $\alpha_{obs}$ is more relevant to star formation. $\alpha_{obs}$ might actually be more meaningful if large gas motions due to the the galaxy potential impede star formation.  However, $\alpha_{obs}$ is a direct measure of the relative strengths of the galaxy potential and self-gravity. We advocate a different indicator of the onset of self-gravitation ($\S$ \ref{sec:forcebalance}), described fully in Paper II.   

\begin{figure*}
\begin{center}
\begin{tabular}{cc}
  \includegraphics[width=0.485\linewidth]{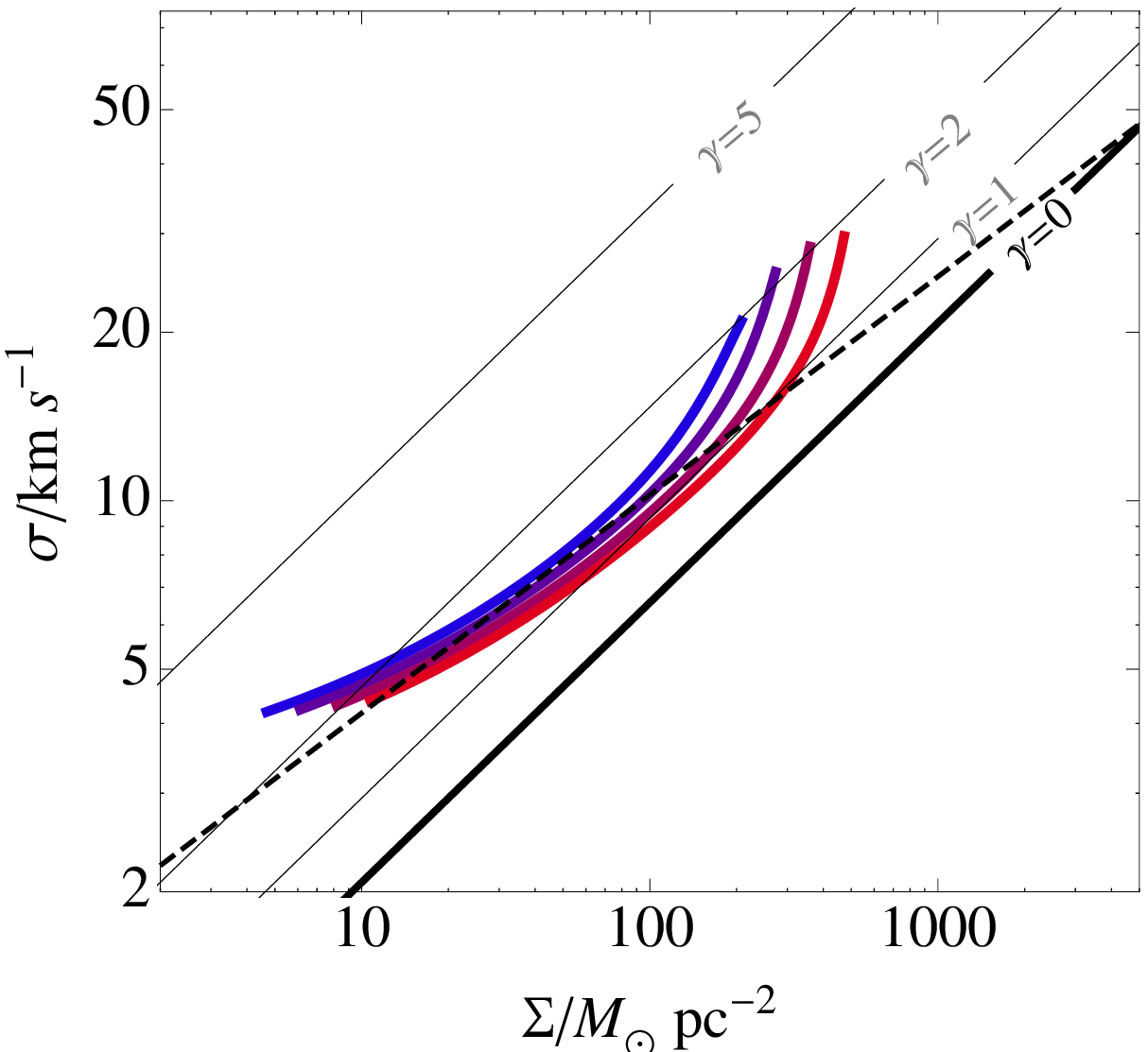}&  \includegraphics[width=0.485\linewidth]{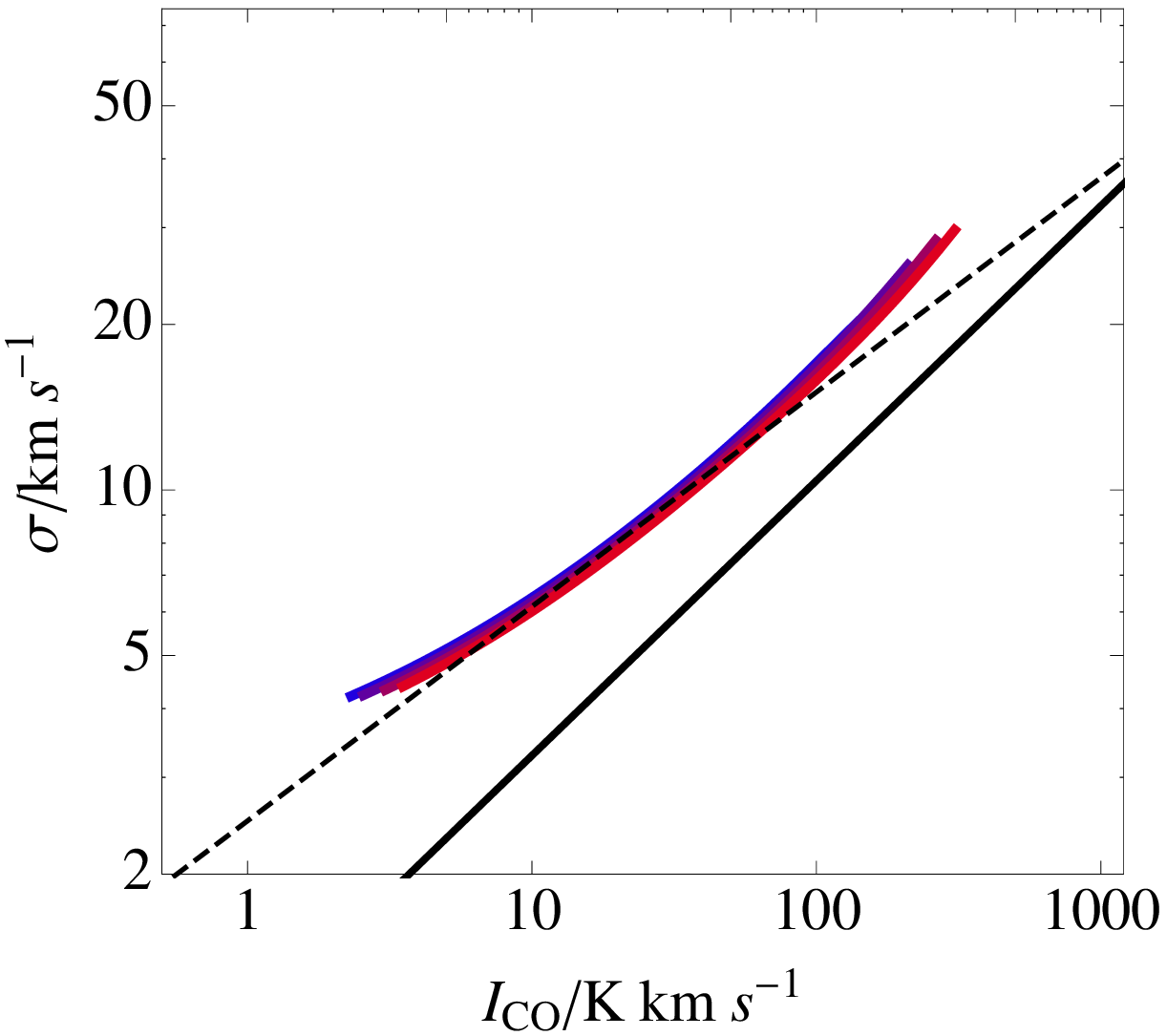}\\
\end{tabular}
 \end{center}
\caption{Predicted trends in $\sigma$ vs. $\Sigma$ (left) and vs. $I_{CO}$ (right) on 150 pc scales throughout the molecular gas disks of nearby galaxies, where $\sigma$ represents the motions required for energy equipartition in the presence of gas self-gravity and the background galaxy gravitational potential.  (Other sources of motion, such as feedback and magnetic fields, have been omitted here for illustration.) Curves track the radial dependence of both quantities, modeled as described in the text, in galaxies with stellar masses in the range 9.25$<$$\log M_{\star}/M_\odot$$<$10.75 in steps of 0.5 $\log M_{\star}/M_\odot$.  
Trends are shown out to 5 $R_e$, to encompass the full extent of the molecular disks of nearby star-forming galaxies \citep{schruba}.   The thick black line marks the relation expected in the case of gas self-gravity (with $\gamma$=0, in the absence of the galaxy gravitational potential) while in the left panel three other gray lines mark an increasing contribution from the external potential (from low to high: $\gamma$=1, 2, and 5).  The black dashed line shows the relation $\log{\sigma}$=0.34 $\log{\Sigma}$ predicted for the net balance of kinetic and (internal and external) potential energies (see $\S$ \ref{sec:netmotions}) at a radius of 2$R_e$ in our empirically-motivated disk model, near the edge of the bright molecular disk.  Energy equipartition in the presence of gas self-gravity and the galaxy potential becomes increasingly recognizable in the main disk at larger galactocentric radius and lower $\Sigma$.  
\vspace*{0.35cm}  
}
\label{fig:sigVSig}
\end{figure*}
\subsubsection{Trends in the average velocity dispersion at fixed size/scale}\label{sec:vdispSigma}

The dynamical state of molecular gas can also be assessed by comparing the ratio of line width, $\sigma_{obs}$, to surface density, $\Sigma_c$ at fixed spatial scale (e.g., \citealt{leroy2016}; Sun et al., in prep.). Gas organized in to clouds in approximate virial equilibrium is expected to follow $\sigma_{obs}$$\propto$$\Sigma_c^{1/2}$ at fixed spatial scale when the scale probes the typical cloud size, $R_c$ (\citealt{heyer}; and see eq. \ref{eq:sigSG}).\footnote{We emphasize that the relation between $\sigma$ and $\Sigma$ at fixed spatial scale is recommended for use as a diagnostic of the dynamical state of clouds when designated specifically by the expression for energy equipartition in eq. (\ref{eq:vdisp}) only when considered on/near the cloud scale.  In other scenarios, the primary factors governing the balance of energies may be different, changing the predicted relation between $\sigma$ and $\Sigma$. On the core scale, for example, the influence of the surrounding cloud material might be best incorporated as the external potential (i.e. \citealt{BP09II}) or, alternatively, as an external pressure (e.g. \citealt{field}). }

Figure \ref{fig:sigVSig} illustrates how $\sigma_{obs}$ due to the combination of self-gravity and the background galaxy potential would depend on $\Sigma_c$. We plot results for the average relation for galaxy disks with several different masses, again using the empirical models in Appendix \ref{sec:appendixGas}. For this example, we adopt a 150~pc beam, typical of extragalactic cloud surveys.\footnote{The predicted velocity dispersions are assumed to probe motion near the cloud scale.  When the observed scale exceeds the typical cloud size, the relevant scale to model is the epicyclic radius, or the scale over which motions due to the galaxy remain coherent, rather than the beam size (see $\S$ \ref{sec:cloudrot} for a discussion of the predicted scales to observe epicyclic motions.}

Averaged across the gas disk, our predicted relationship between $\sigma_{obs}$ and $\Sigma_c$ appears shifted to slightly higher values than in the pure self-gravitating case, and with more curvature. The largest systematic deviations occur at large $R_{gal}$.  Given the lower $\Sigma_c$ assumed there, our model predicts a weakening of the gas self-gravity relative to the galaxy potential.  As a result, $\sigma_{obs}$ traces a shallower dependence on $\Sigma_c$. Such elevated line widths and a deviation from the expected self-gravity scaling could serve as an observational indicator that that motions due to the galaxy potential are playing an dominant role in setting $\sigma_{obs}$.  

We can describe how the trend emerges by expanding both terms in eq. (\ref{eq:vdisp}) to first order around some multiple of $R_e$.   
At radius $R$=$nR_e$, we now write the 1st term (varying like $e^{-R}$) as
\begin{equation}
\sigma_{sg}^2=\left[\pi G R_c \Sigma_0 \left(e^{-n}(1+n)\right)\right]\left\{1-\frac{1}{1+n}\frac{R}{R_e}\right\}\label{eq:sgexp}
\end{equation}
and the second term, which varies like $R^{-\eta}$, as
\begin{eqnarray}
\sigma_{gal}^2=\left[\frac{V_c^2 R_c^2}{C}\left(\frac{(1+\eta)}{R_e^2n^2}\right)\right]\left\{1-\frac{\eta}{(1+\eta)n}\frac{R}{R_e}\right\}\label{eq:galexp}
\end{eqnarray}
Here we let $\eta$ (and the constant factor $C$) reflect the different radial trends predicted in the case of either isotropic or non-isotropic motion considered in the model.  For isotropic motions, $\sigma_{gal}^2\sim(\kappa R_c)^2\approx2(V_c R_c)^2 R^{-2}$ in the flat part of the rotation curve so here $\eta=2$ and $C$=1.  For non-isotropic motions $\sigma_{gal}^2$ is dominated by vertical motions $(\nu R_c)^2\sim(\kappa R_c)^2 R (2z_0)^{-1}\sim R^{-1}$, so $\eta$=1 and $C=(2z_0)$.

Over most of the disk, the coefficients in front of the bracketed terms in expressions (\ref{eq:sgexp}) and (\ref{eq:galexp}) above are comparable, as suggested by Figure \ref{fig:paramspaceMW}.  
As a result, 
the two bracketed terms combine to approximately yield a single dependence 
\begin{equation}
\sigma_{sg}^2+\sigma_{gal}^2\approx 1-\mathcal{F}\frac{R}{R_e}
\end{equation} 
which is the first order approximation to the exponential $e^{-\mathcal{F}R/R_e}$ where 
\begin{equation}
\mathcal{F}=\frac{\eta+n+2\eta n}{(1+n)(1+\eta)n}.
\end{equation} 
Thus, moving towards large galactocentric radius the velocity dispersion $\sigma=(\sigma_{sg}^2+\sigma_{gal}^2)^{1/2}$ varies approximately as $\Sigma^{5/(6n)}$ for large $n$ in the isotropic case.  

According to our model, the relation between $\sigma$ and $\Sigma$ will vary across several key environments.  \\

\noindent\underline{Galaxy disks} \\
Since self-gravity and the external potential are of similar strength across the bulk of typical galaxy disks, we expect cloud-scale motions in the majority of the molecular gas to be well-described by the approximation given above.  Particularly in the zone between $R_{gal}$=0.5 and 2$R_e$ (spanned by the brightest molecular emission; \citealt{schruba}), the slope of the relation between $\log{\sigma}$ and $\log{\Sigma}$ modeled in Figure \ref{fig:sigVSig} decreases from 0.5 to $\approx$0.3.  At these radii galaxy rotation curves have mostly flattened out and so the trend is nearly independent of galaxy mass.   A slightly shallower slope is expected in the case of non-isotropic galaxy-induced motions, although these mostly remain comparable to the estimated isotropic motions across this zone.  \\

\noindent\underline{Galaxy centers} \\
In the centers of galaxies where gas surface densities are high but the galaxy potential still dominates self-gravity, the dependence of $\sigma$ on $\Sigma$ is much steeper.  Here we expect the gradient in the potential to be larger in the plane than in the vertical direction, so motions in both the isotropic and non-isotropic cases will vary approximately as $\kappa R_c$, leading to $\sigma$$\approx$$\Sigma^{0.25 n^{-1}}$, which becomes super-linear at radii less than $R\approx0.25R_e$.  According to the strong variation in the CO-to-$H_2$ conversion factor in galaxy centers predicted by the model (discussed later in $\S$ \ref{sec:xco}), however, $\sigma$ is predicted to show a less strong dependence on the CO line intensity $I_{CO}$ at inner radii than with $\Sigma$ as illustrated in the right panel of Figure \ref{fig:sigVSig}. \\

\noindent\underline{Bars and spiral arms}\\
With the addition of dynamical features, galaxies may show that $\sigma$ is locally raised even further above the self-gravity expectation than indicated by the axisymmetric models assumed in Figure \ref{fig:sigVSig}.  But on average, in the strongest cases this should be limited to the innermost radii or well beyond the edge of the bright molecular disk, given the dynamical characteristics of nearby star-forming disks.  In the case of bars, deviations are expected to be largest at small galactocentric radius and decrease towards the bar corotation radius $R_{CR,bar}$, where the bar rotates with the same angular rate as the disk material.  This should occur at or beyond the edge of the brightest molecular emission, given that the corotation radii of bars typically fall in the range $R_{CR,bar}$=0.3-0.6 $R_{25}$=1.5-3$R_e$=0.4-0.5 $R_{90}$ (\citealt{raut}; with $R_e$, $R_{50}$ and $R_{90}$ as measured for nearby molecular disks by \citealt{schruba}). Spiral corotation radii are located even further out (e.g. \citealt{elmea89}).  \\

\noindent\underline{Galaxy outskirts} \\
Moving toward galaxy outskirts (not shown) 
the gas surface density becomes so low that the galaxy potential induced motions dominate those due to self-gravity, leading to large `super-virial' motions and considerable offsets from the self-gravity line portrayed in Figure \ref{fig:sigVSig}.  
At large enough radius where the background potential becomes less disk-like and more triaxial (dominated by dark matter) the super-virial gas motions on cloud scales are expected to be primarily isotropic.  Here the one-dimensional velocity dispersion $(\nu h)^2=4\pi G \rho h^2$, where $h$ is the gas scale height or cloud radius and $\rho$ is the local (dark matter) density.  

Note that clouds at these large radii, which have been observed in the MW to be very small and sustain much larger velocity dispersions at fixed size than those held together by gravity alone, are thought to be pressure-confined (see \citealt{elmegreen89} and \citealt{field}).  Our model provides an equivalent explanation, but rather than invoke pressure, i.e. balancing the weight of the background galaxy, we relate gas motions directly to the galaxy potential itself.  

\subsection{A first comparison to observations}\label{sec:datacomp}
\begin{figure*}[t]
\begin{center}
\begin{tabular}{cc}
\includegraphics[width=.415\linewidth]{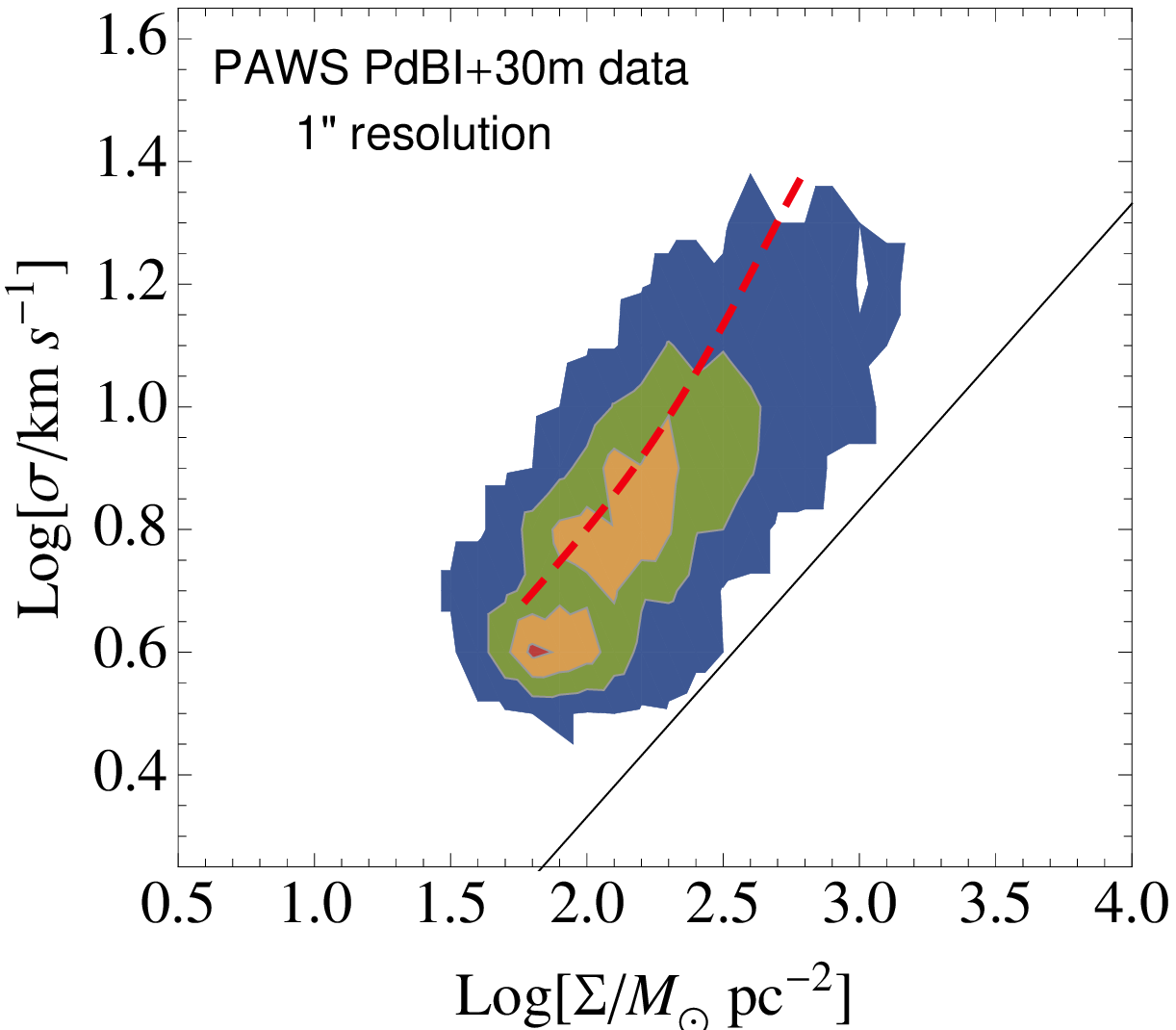}&
\includegraphics[width=.415\linewidth]{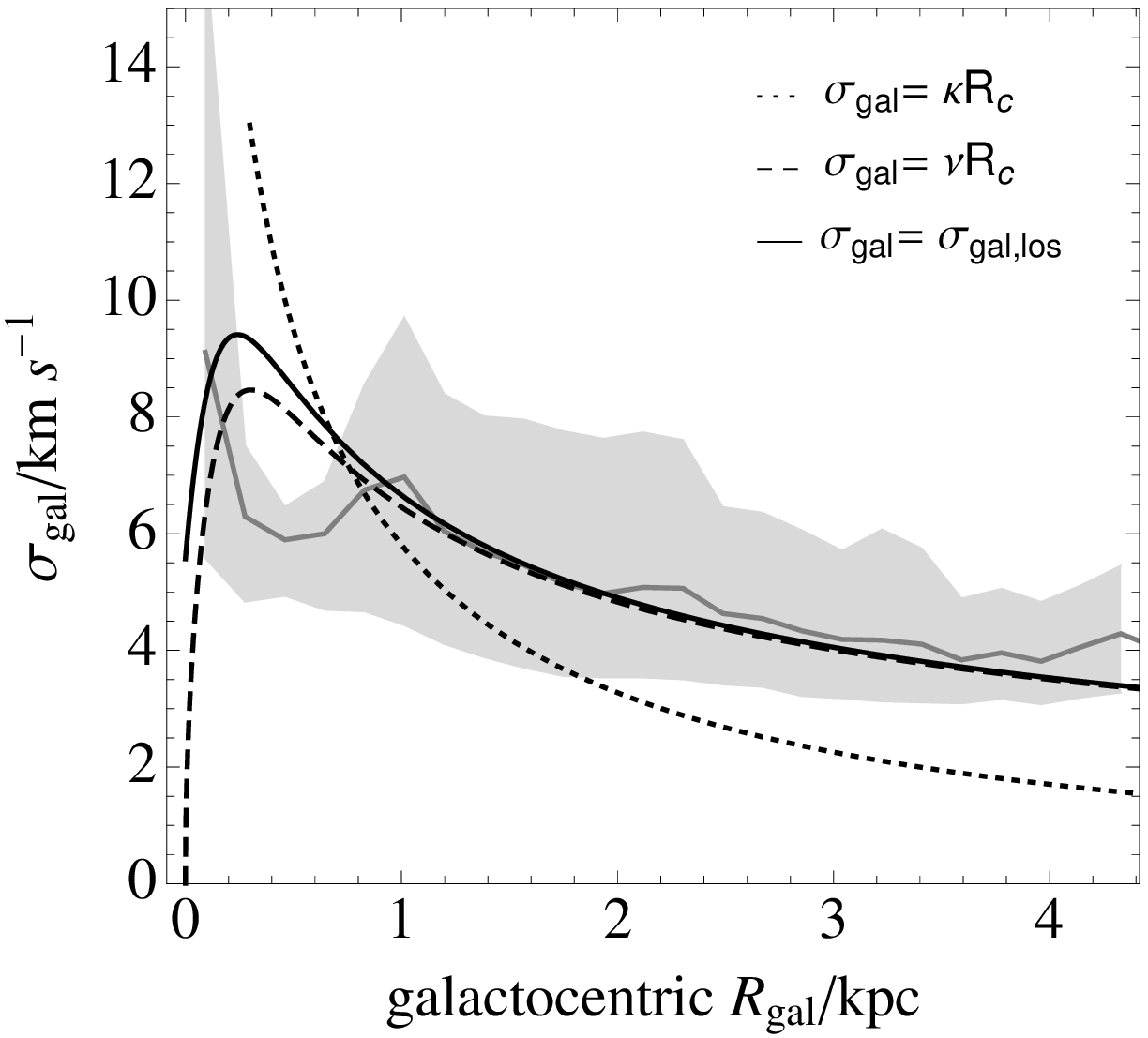}\\
\includegraphics[width=.415\linewidth]{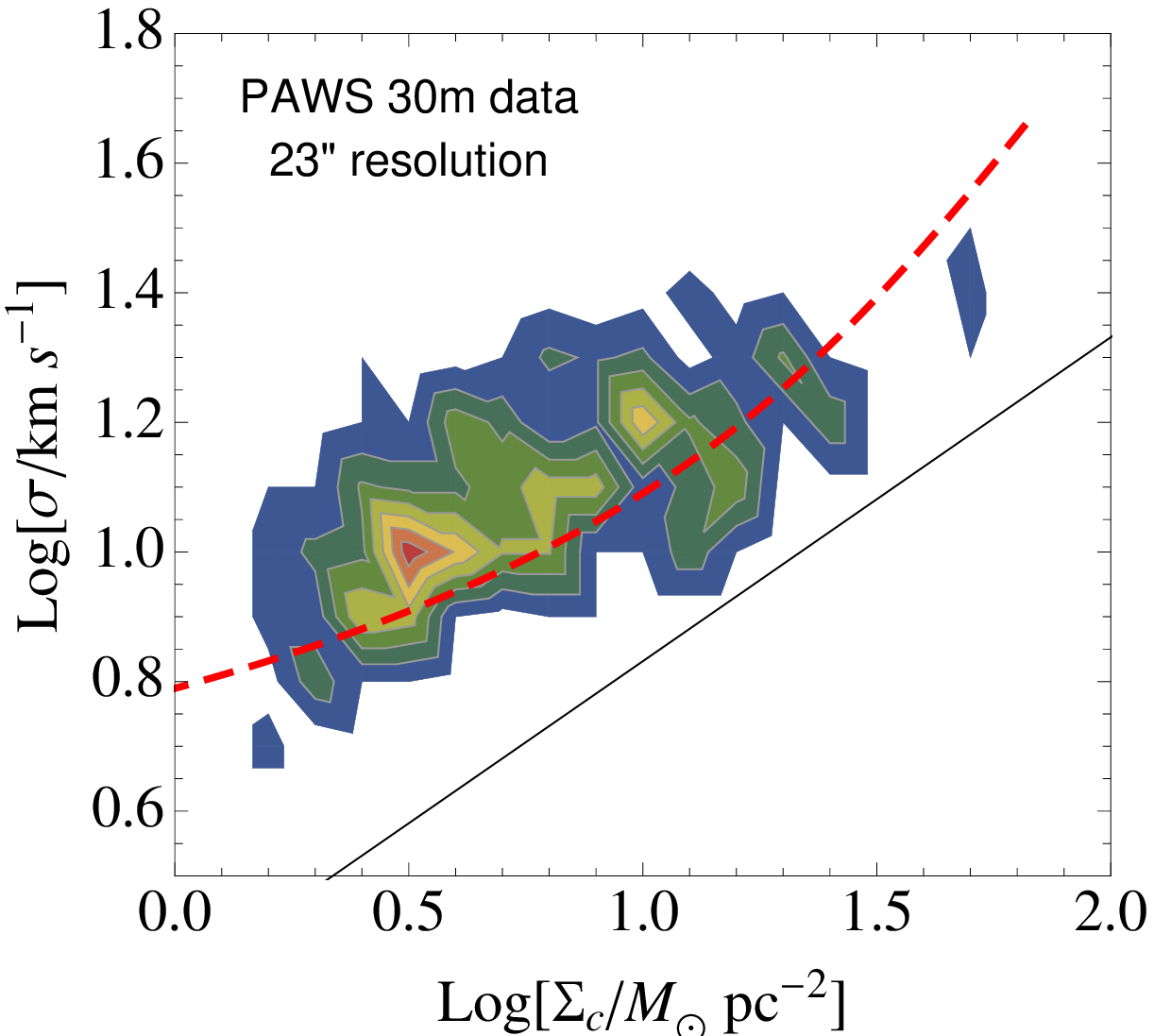}&
\includegraphics[width=.415\linewidth]{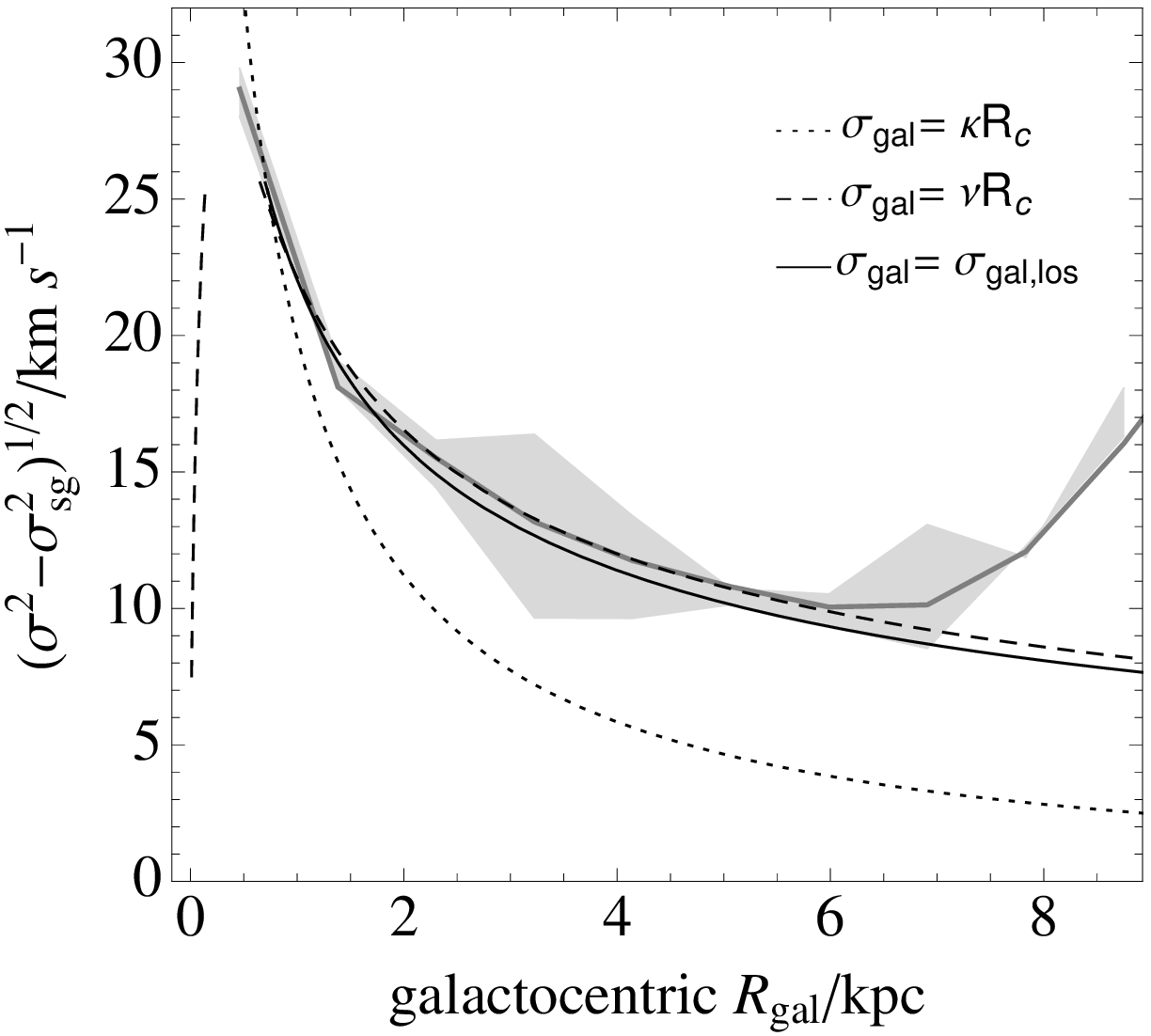}
\end{tabular}
\end{center}
\caption{Comparison of observed and modeled motions in the molecular gas of M51 at 1"=40pc resolution across the PAWS field of view (top row) and derived from the PAWS (single-dish) 30m observations of CO(1-0) at 23" resolution (bottom row).  (Left) Contours of the observed velocity dispersion $\sigma$ vs. gas surface density $\Sigma$.  The black line shows the prediction for motions due to self-gravity $\sigma_{sg}$ (top) and 5$\times\sigma_{sg}$ (bottom).  The red dashed line indicates the trend predicted by our model in a galaxy with the mass of M51, across radii matched to the PAWS field of view (top) or across the entire disk (bottom), in the latter case with an arbitrary assumed filling factor of 0.1 (clumping factor $c$=10).  (Right) The gray shaded area highlights the range in the 1D super-virial motions $(\sigma^2-\sigma_{sg}^2)^{1/2}$ throughout the molecular disk measured by subtracting from the observed $\sigma$ an estimate for $\sigma_{sg}$ defined in relation to the observed $\Sigma$ (see text).  The azimuthal average at all $R_{gal}$ is indicated by the solid dark gray line.  Black curves show model predictions adopting $R_c$=25 pc for the PAWS data.  The 30m data are matched assuming $R_c$=80 pc. The black solid line shows the model prediction for $\sigma_{gal,los}$ due to the galaxy potential in the non-isotropic velocity case.  The black dotted line shows the prediction assuming isotropic motions in 1D estimated as $\kappa R_c$.  The black dashed line represents $\nu R_c$ in the non-isotropic model.  \vspace*{0.35cm}  }
\label{fig:m51prof}
\end{figure*}

\begin{figure*}
\begin{center}
\begin{tabular}{cc}
 \includegraphics[width=0.475\linewidth]{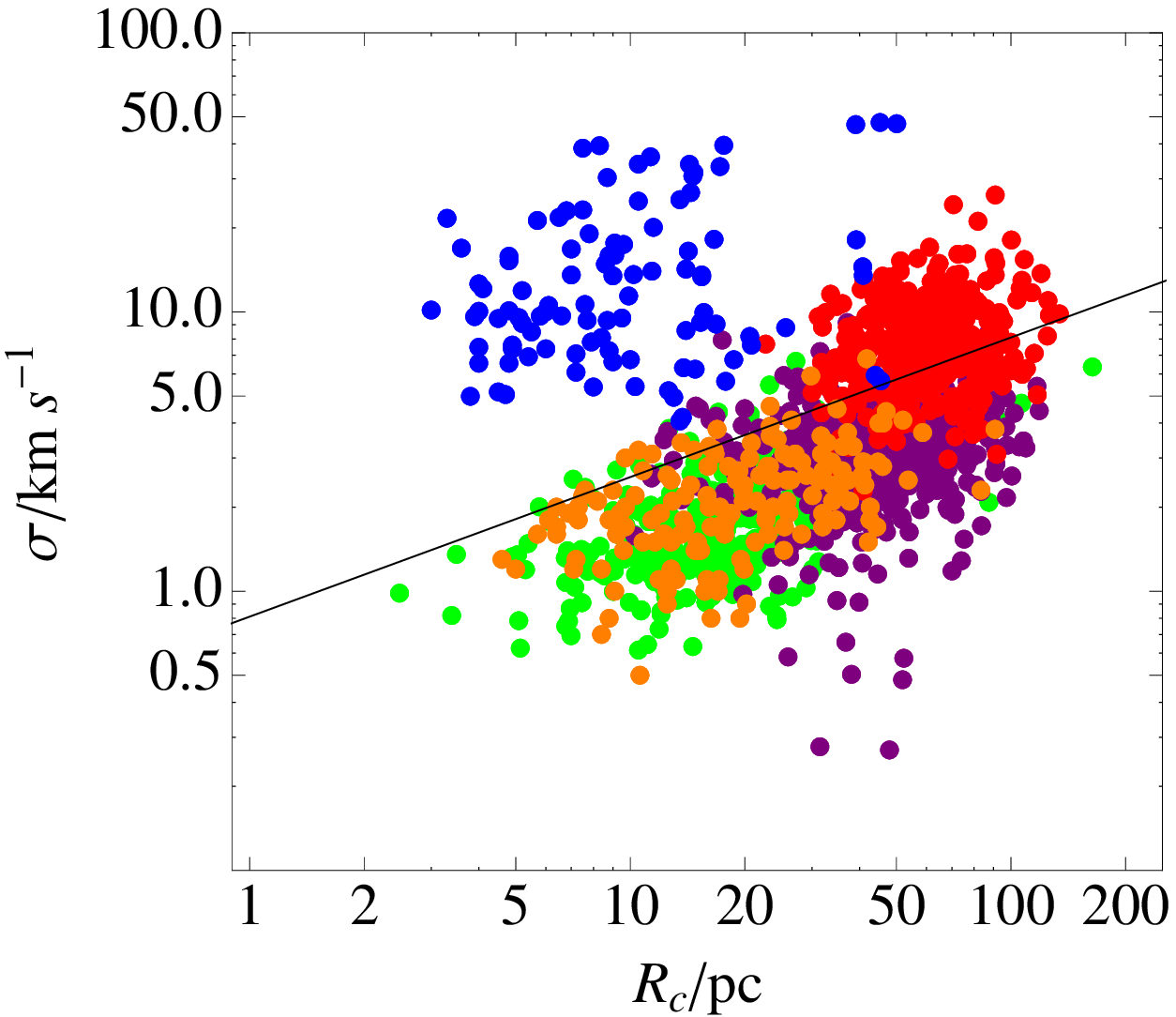}\vspace*{.15cm}&  \includegraphics[width=0.465\linewidth]{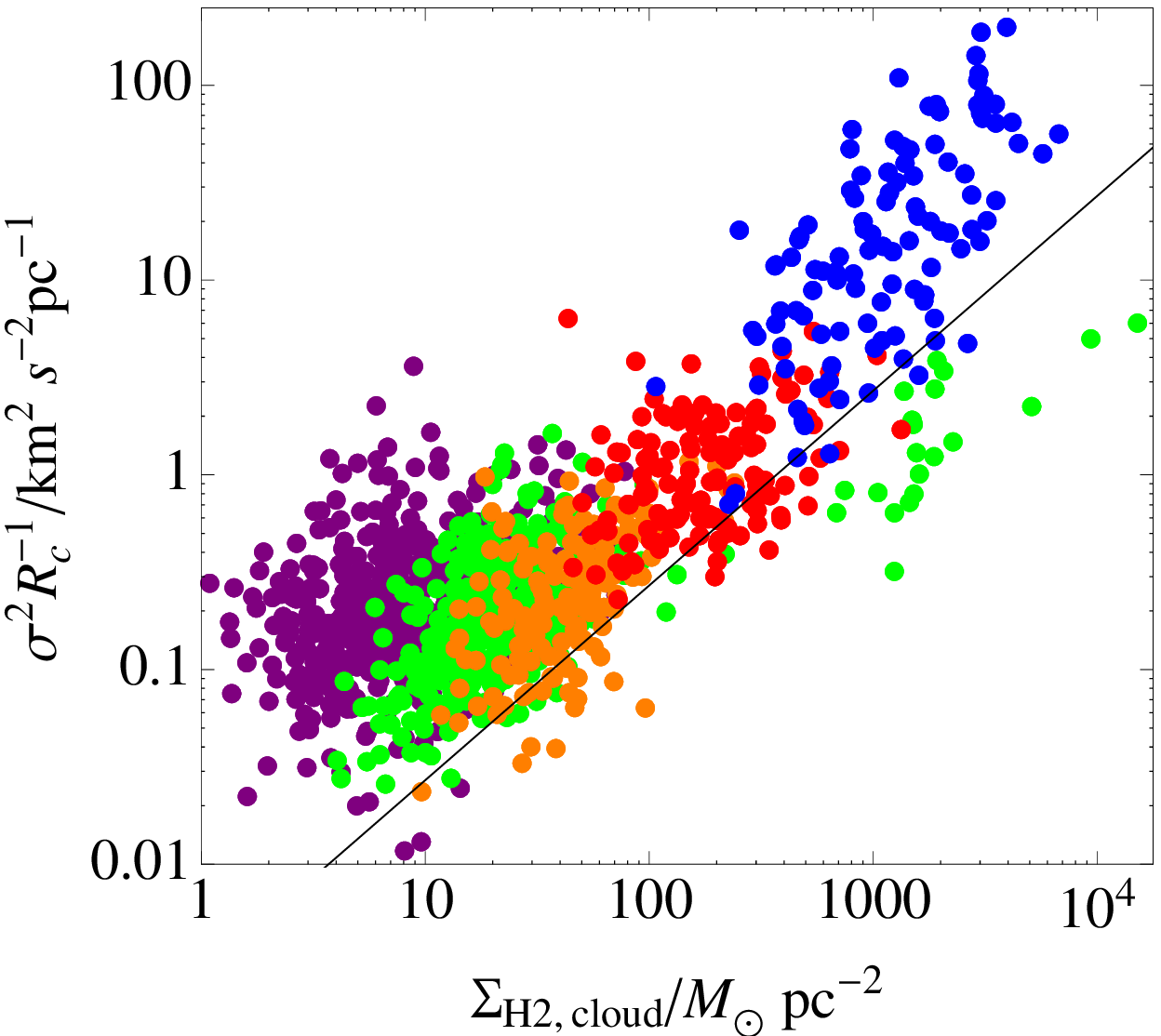}\\
  \includegraphics[width=0.475\linewidth]{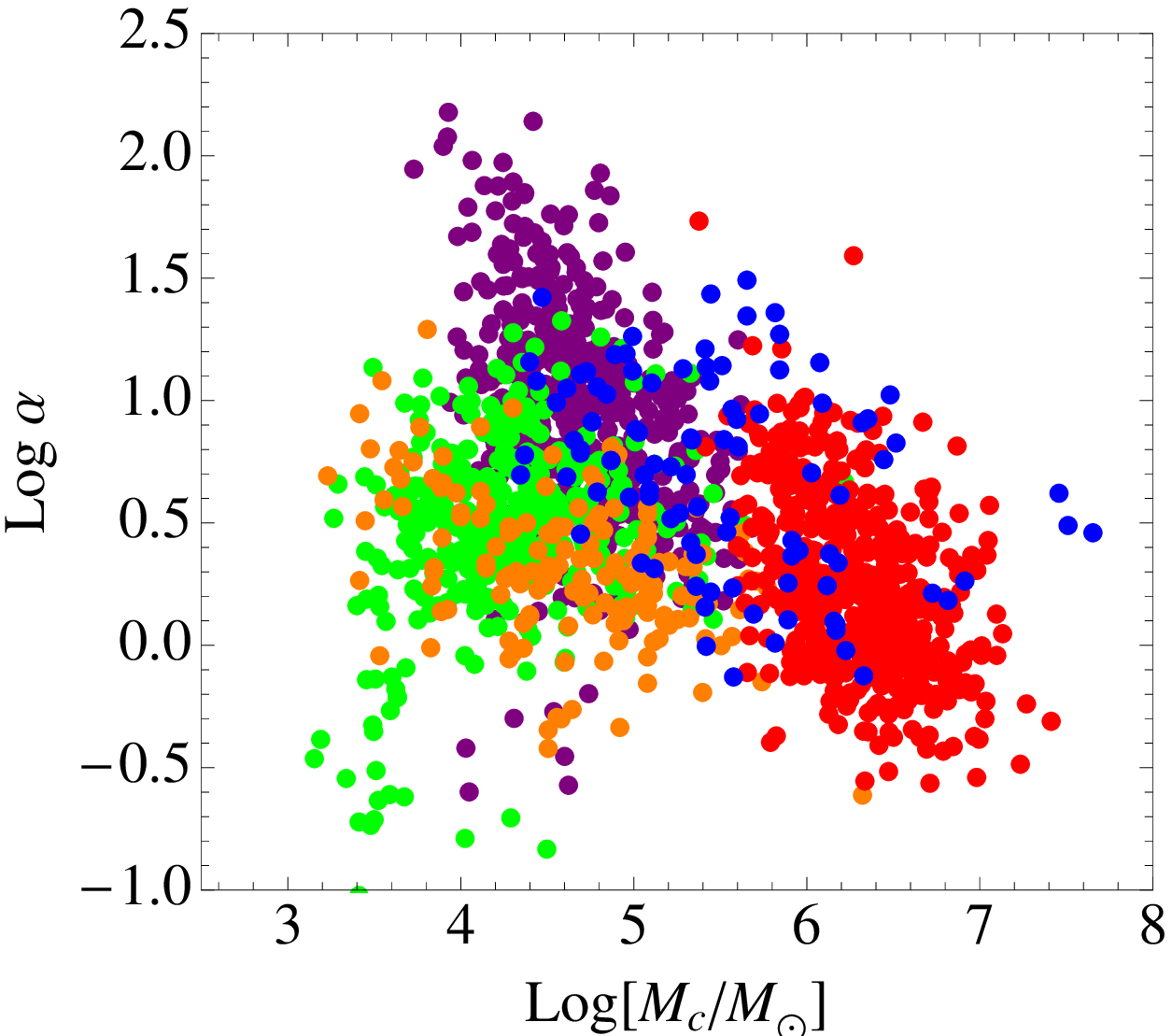}&  
  \includegraphics[width=0.475\linewidth]{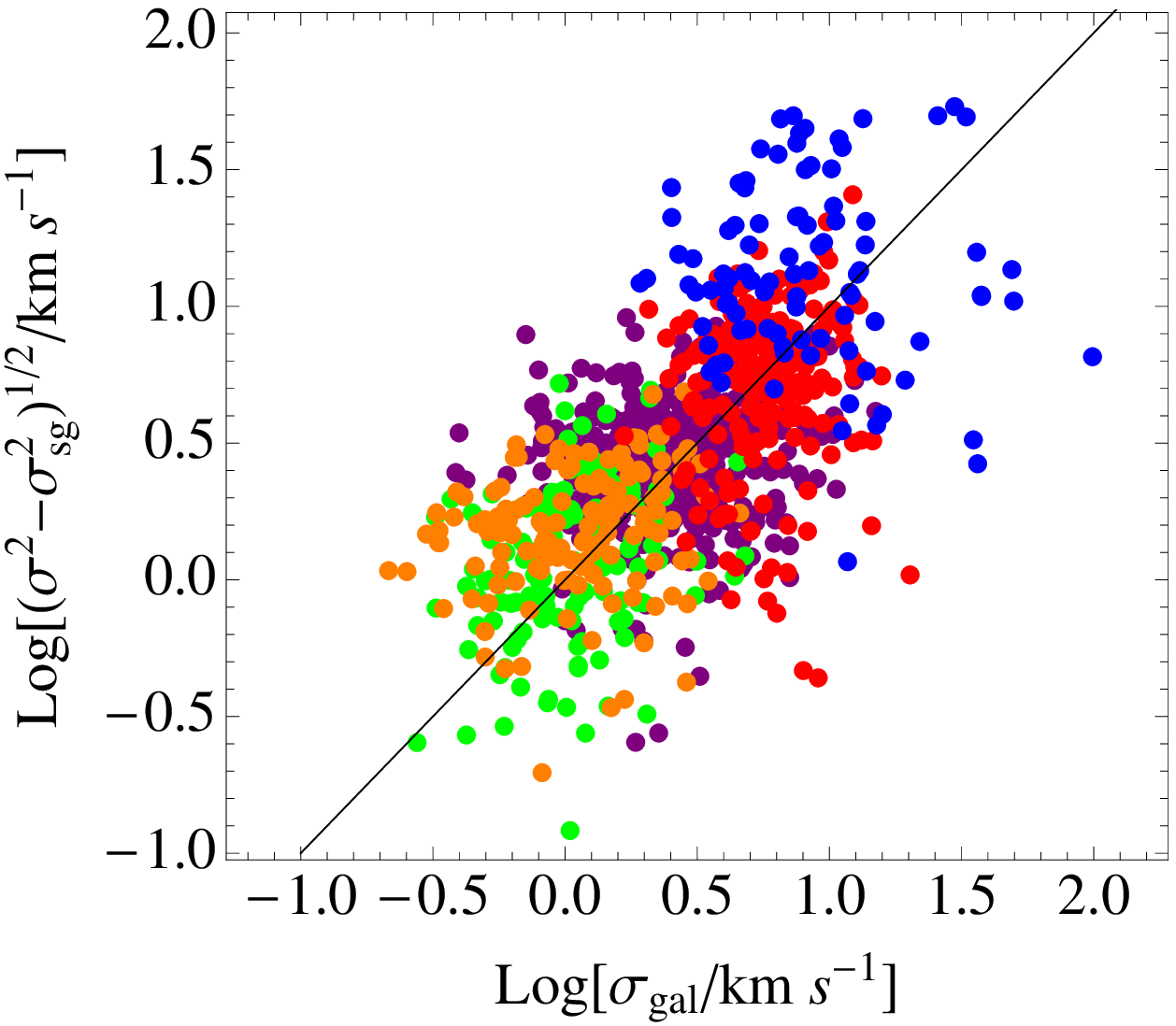}
 \end{tabular}
 \end{center}
\caption{Molecular cloud properties in M51 (red), M33 (purple), the LMC (green), the MW disk (orange) and center (blue).  Each point represents an individual cloud.   The top left panel shows the observed velocity dispersion $\sigma$ vs. the measured cloud size $R_c$.  The black line indicates the size-linewidth relation measured by \cite{solomon} in a sample of Solar Neighborhood clouds.  The top right shows the 'size-linewidth relation coefficient' $a$=$\sigma^2/R_c$ vs. cloud surface density $\Sigma_c$.  The trend predicted for clouds in virial equilibrium is indicated by the black line.  The balance between internal kinetic and potential energies is shown in the bottom left panel.  For clouds in virial equilibrium $\alpha$=1 while $\alpha$=2 for bound clouds.  The bottom right panel shows the relation between super-virial motions $\sigma^2-\sigma_{sg}^2$ (i.e. those beyond what is expected due to the cloud's self-gravity $\sigma_{sg}$) and the motions $\sigma_{gal}$ predicted by our model of 3D gas motions driven by the background galaxy potential on the scale of each cloud. 
\vspace*{0.35cm}  }
\label{fig:catalogscomp}
\end{figure*}
Here we present a preliminary comparison of the model with observations. First we consider the basic trends in measurable gas properties expected within a single galaxy. Then we examine how well the model describes differences that emerge between cloud populations depending on global galaxy properties.  

\subsubsection{Variations in the dynamical state of the gas in M51 measured on cloud scales}

The PAWS survey \citep{schinnerer} of CO~(1-0) emission in M51 yielded one of the first cloud-scale views of molecular gas across a grand-design spiral galaxy. Here we compare the molecular gas velocity dispersion measured from PAWS to the models developed above. 

We take $\sigma_{obs}$ and $\Sigma$ from the PAWS zeroth and second moment maps \citep{pety} at their native 1" resolution and at the $23\arcsec$ resolution of the PAWS single dish survey. These correspond to 40 and $\sim 1$~kpc at the adopted $D$=7.6 Mpc distance to M51. PAWS includes short and zero spacing data, and the ISM of M51 is molecule-dominated over the PAWS field \citep{hitschfeld}. Therefore these maps should accurately capture the gas distribution. The single dish map, though low enough resolution that it does not sample cloud scales, covers a wider area than the interferometer map.

We estimate the epicyclic frequency, $\kappa$, from the PAWS rotation curve derived by \citealt{meidt}. For the non-isotropic case, we estimate $\nu$ using the scaling relations to link the stellar mass of M51 with a model for the stellar mass surface density and stellar scaleheight (see Appendix \ref{sec:appendixnu}). This estimate of $\nu$ is more approximate than our measurement of $\kappa$.

The left panel of Figure \ref{fig:m51prof} shows $\sigma_{obs}$ vs. $\Sigma$ at the fixed 40 pc (top) and 1~kpc (bottom) scales.  In both cases motions exceed those predicted for self-gravity alone (illustrated by the black line). The data appear well-approximated by our models that combine self-gravity and galactic forces, shown in red.  These red model curves are the same predictions shown in Figure \ref{fig:sigVSig} for a galaxy with the mass of M51, extending over sthe appropriate field of view for each data set.  To approximate the effect of resolution on the surface density observed at  the lower resolution of the single dish (30m) data, the curve on the bottom is shifted to lower surface densities than in our nominal cloud-scale surface density model by an arbirtrary factor of 10 (cloud-scale  clumping factor of 10) chosen to match the observations.

The right panels further illustrate the good match between our model and the PAWS observation. There we show the radial variation in the component of observed motions not due to gas self-gravity, i.e., $\sigma_{gal}^2 = \sigma_{obs}^2-\sigma_{sg}^2)^{1/2}$ with $\sigma_{sg}$ estimated from the observed 
surface density.

In both the top and bottom rows, the motions in excess of self-gravity (gray band) match the predictions from our model well.  In particular, our data appear well-matched to the non-isotropic case, in which vertical motions $\nu R_c$ are larger than predicted in the isotropic case where they are $\kappa R_c$.  Given the nearly face-on inclination of M51, vertical motions are expected to dominate the observed line-of-sight velocity dispersion. 
For completeness, the solid line shows the prediction for the full non-isotropic model that includes the projection of in-plane motions $\kappa R_b\sin{i}$ along the line of sight, with $R_b$ set by the beam size.

The 30m data shows a similar result, with the observed line widths well-described by the vertical term in the anisotropic case. The measured velocity dispersions are much higher than in the PAWS map. This likely reflects a mixture of blurring of the rotation curve (``beam smearing'') and the greater sensitivity of the 30m data to spatially-extended emission. The slight increase in $\sigma$ toward the outermost radii may be due to increasingly low signal-to-noise at the map edge.

The cloud scale $R_c$ that must be assumed in order to generate the comparison models are chosen to be $R_c = 25$~pc for the cloud-scale data and $80$~pc for the single dish data. These values are chosen to match the observed line widths, but they also match the measured cloud sizes in M51 within the uncertainties. \citet{colombo2014a} found a median rms cloud radius of 32~pc (corresponding to an rms size of 64~pc), characterizing $\sim 50\%$ of the emission. \cite{pety} estimated a scale height of the gas of $\approx 100$~pc using calculations similar to those we carry out here.

\subsubsection{Variations from galaxy to galaxy}

M51 shows clear signatures of motions driven by the galaxy potential. Next we apply our model to interpret differences among the cloud populations hosted by galaxies with different stellar masses and orientations. 

For this comparison, we use catalogs of cloud properties estimated from CO emission in four galaxies: M51 \citep{colombo2014a}, M33 \citep{druard}, the LMC \citep{hughesLMC} and the MW. We further separate the Milky Way into disk \citep{heyer} and center \citep{oka} populations.  We use the properties as catalogued, only adjusting where needed to calculate a set of uniformly defined cloud masses, rms velocity dispersions, and rms sizes. We note that comparisons between cloud populations extracted with varying decomposition strategies, applied to observations with non-uniform spatial and spectral resolutions, should be interpreted with caution \citep{hughesI}. Table \ref{tab:galprops} reports the key properties of each target.

Figure \ref{fig:catalogscomp} shows the properties of clouds from the four galaxies in several key parameter spaces.  In the top left, we see that clouds in M51 have larger line widths at fixed size than those in the lower-mass M33 and LMC, as previously noted by \cite{hughesI}.  As shown by \citet{oka}, Galactic center clouds also have higher velocity dispersions than their counterparts in the disk of the MW. 

Clouds also populate different regions in plots at the top right and bottom left of Figure \ref{fig:catalogscomp}. The top right panel shows the 'line-width size relation coefficient' $a$=$\sigma^2/R_c$ vs. cloud surface density $\Sigma_c$. The bottom left shows virial parameters vs. cloud mass $M_c$. These plots show many of the same characteristics seen in Figures \ref{fig:cloudReln} and \ref{fig:alpha}. The rise in $\alpha_{obs}$ at low cloud mass, and the curvature of the coefficient of the line width size relation as a function of surface density suggest that clouds across these galaxies indeed resemble our model calculations.

Differing cloud demographics and rotation curve shapes preclude a single model prediction that can be plotted in the first three panels. We can note some general trends, however. 
First, clouds arising in regions where the galaxy potential is
relatively weak (either because the gas surface density and self-gravity is particularly high or the potential itself only weakly varying; e.g. LMC, Solar Neighborhood, M51), show better proximity to
virial equilbrium (alpha=1) than clouds in the galaxy center or M33.  
Also, the strong offset of Galactic center clouds towards much larger normalized linewidths than expected for their surface density is reminiscent of the offset portrayed in Figure \ref{fig:sigVSig} at inner radii (at high gas surface density), where the galaxy potential dominates self-gravity.

In the bottom right panel of Figure \ref{fig:catalogscomp}, we directly compare the cloud catalogs to our model. To do this, we plot the observed 'super-virial' motions against the predicted motions induced by the galaxy in our model. As above, here the super-virial motions refer to the difference in quadrature between $\sigma_{obs}$ and the dispersion predicted from only self-gravity. Except in the case of M51, we estimate $\kappa$ for each cloud based on its present galactocentric radius and the rotation curve estimated based on the mass of its host galaxy. For the Galactic center clouds, positions are inferred from their LSR velocities, also adopting a MW-mass rotation curve. Likewise, for each cloud we assign the vertical frequency $\nu$ based on the stellar density inferred at the cloud's position, again using galaxy scaling relations to estimate the stellar surface density and scale height.

Despite the approximate nature of the model, the bottom right panel of Figure \ref{fig:catalogscomp} display remarkable agreement between the predicted and observed super-virial motions. Although scatter remains, there is much less systematic deviation from galaxy to galaxy than any in any of the other parameter spaces shown in the Figure.

Here we assume that the motions are anisotropic in a spherical region of radius $R_c$, which we set to the measured cloud size.  This choice appears to provide the best match to the observations, particularly with regard to the clouds in the more face-on galaxies M51 and M33.  As we found for M51 in the previous section, the vertical component that dominates the line-of-sight at low $i$ exceeds the modeled $\kappa R_c$ but can be well described by $\nu R_c$.

At the inclination of M33, the line-of-sight contains motions that arise roughly half from the vertical direction and half from in-plane motions. We find that both are necessary to match the observations. For galactic center clouds, on the other hand, the line-of-sight records motions in the plane and we find that motions in the plane $\kappa R_c$ provides a good match to the observations.

Overall, the bottom right of Figure \ref{fig:catalogscomp} suggests that the forces exerted by the external galaxy potential may offer a compelling interpretation for observed cloud-scale gas motions.  Accounting for the galaxy potential tends to unify clouds across different environments, and even Galactic center clouds resemble those in less extreme environments. 
 
 \begin{table}
\begin{center}
\caption{Properties of cloud host galaxies}\label{tab:galprops}
\begin{threeparttable}
\begin{tabular}{lccc}
Name&Distance\tnote{1}&inclination\tnote{2}
& log stellar mass\tnote{3}\\

& [Mpc] & [degrees] (Ref)& [$M_{\odot}$] (Ref)\\
\tableline
 M51&7.6&21(1)&$10.6(1)$\\
 M33& 0.84&56(2)&9.7(2)\\
 LMC&0.05&35(3)&9.3(3)\\
 MW& -- & -- & 10.7(4)
 \end{tabular}
 \begin{tablenotes}
\item[1]All distances adopted from \cite{hughesI}.
\item[2]References for inclinations: (1) \cite{colombo14b}, (2) \cite{paturel}, (3) \cite{vdM}.
\item[3]References for stellar masses: (1) \cite{leroy08}, (2) \cite{corbelli}, (3) \cite{kim98}, (4) \cite{licquia}. 
 \end{tablenotes}
\end{threeparttable}
 \end{center}
 \end{table}

\section{Further Predictions of the Model}
\label{sec:predictions}

Despite the simplicity of the model described above, it appears to match observations well. As we describe in this section, there are several other observational tests that could be used to assess the validity of the model and the importance of the local galaxy potential.  We highlight signatures of rotation, virial equilibrium and cloud morphology.  

\subsection{Observational signatures of cloud rotation}\label{sec:cloudrot}

Coherent motions on the scale of clouds can be directly observed. If the motions described above are present and dominant, then rotational motion about the galaxy center should be visible at or even below the scale of a cloud.  These motions should appear as a measurable velocity gradient across the cloud in the same sense as the local bulk velocity field.  A strong first-order test of our model is thus to search for signatures of cloud-scale velocity gradients aligned with the large scale velocity field.  

Intrinsically elliptical cloud shapes, such as expected according to the discussion in $\S$ \ref{sec:morphs}, will introduce a misalignment between the observed orientation of the velocity gradient and the true rotational line-of-nodes. Following the treatment applied to elliptical stellar orbits, we can expect the misalignment to depend on the cloud axis ratio, the galaxy inclination, and viewing angle (\citealt{teuben}; \citealt{franx}; \citealt{wong}).    

Several other factors may obscure the predicted motions.  As showed by \cite{burkert}, the long-wavelength modes of isotropic turbulence produce a net projected velocity gradient across clouds, imitating the signatures of rotation.  Another exception is when the whole cloud has some spin angular momentum, i.e. inherited during the formation process (e.g. \citealt{ros03}; \citealt{dobbs08}), or as a result of interaction with other clouds, whose potentials we currently ignore).  In this case, the observed kinematics may reflect the combined epicyclic motion and the cloud's rotation.  

An evaluation of how well the predicted velocity gradients compare with the observations to date is warranted, but presently outside the scope of this work.  We note, though, that the observed velocity gradients are  roughly consistent with the epicyclic motions in our model.  For the clouds in M33, \cite{ros03} measure an average (deprojected) gradient $\approx$0.05 km s$^{-1}$ pc$^{-1}$. This is comparable to the average $\kappa$ expected at 1$R_e$ for a galaxy of its stellar mass (see Table \ref{tab:galprops}).  Cloud-scale velocity gradients in the LMC are similar to those in M33 and the MW (Hughes; PhD thesis), and imply that in the LMC cloud rotation may be related to the kinematics of the surrounding gas.  Based on a high resolution study of the molecular gas in early type galaxy NGC 4526, \cite{utomo} suggest that cloud rotation is due to shear and find that observed gradients are consistent with the local rotation.  

\subsection{Implications for the CO-to-H$_2$ conversion factor}\label{sec:xco}
\begin{figure}[t]
\includegraphics[width=0.975\linewidth]{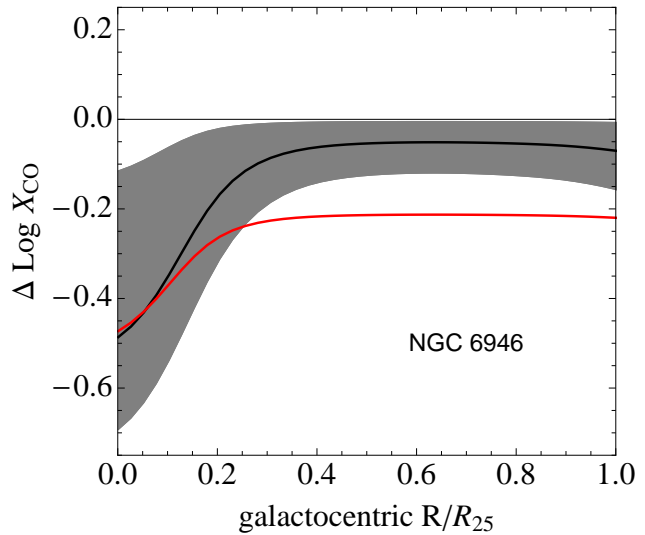}
\caption{Example of the difference in CO-to-H$_2$ conversion factors $X_{CO}$ estimated with and without taking into account the galactic potential in the virial energy balance of molecular clouds.  The galaxy potential is modeled after the properties of the nearby barred galaxy NGC 6946 (see text).  The black curve shows the variation expected for cloud with radius $R_c$=30 pc, in which gravitationally-induced motions driven by the host galaxy are isotropic on the cloud scale.  The gray band indicates the range expected for cloud sizes between 15 pc (top edge) and 50 pc (bottom edge).  The red curve illustrates the change to $X_{CO}$ predicted in the case of non-isotropic cloud-scale motions due to the galaxy.  
\vspace*{0.35cm}  }
\label{fig:xco}
\end{figure}

Contrasting the CO luminosity and dynamical masses $M_{dyn}$ of molecular clouds offers important constraints on the CO-to-H$_2$ conversion factor $\alpha_{CO}$ (\citealt{solomon}; \citealt{scoville}; \citealt{dame}; \citealt{bolatto}; see \citealt{bolatto2013} for a review). Often the dynamical mass is determined based on the assumption that the internal kinetic energy balances the cloud's potential energy but neglecting the potential of the galaxy. In our model, the host galaxy potential should affect dynamical mass measurements everywhere that we predicted super-virial line widths above.

Using $\sigma_{obs}$ to estimate a virial mass $M_{dyn}$ will overestimate the true molecular gas mass $M_{mol}$ when $\kappa R_c$$\gtrsim$$\sigma_{sg}$. Consequently, measurements of $\alpha_{CO}$=$M_{mol}/L_{CO}$ based on these estimates will also be high. The magnitude of the overestimate will be $\kappa^2 R_c^2/\sigma_{sg}^2(1+\sin{i}^2)$=$5\kappa^2 R_c/G\Sigma_c (1+\sin{i}^2)$. Here the $\sin{i}$ term represents the additional contribution to the line-of-sight velocity dispersion from beam-smearing.

The magnitude of the overestimation should be largest in galaxy centers or bars, where $\kappa R_c/\sigma$$>$$>$1. The effect should decrease toward galaxy outskirts. Our model predicts little bias in measurements of $\alpha_{CO}$ from the Solar Neighborhood situated in the outskirts of the MW, for example.

The true value of $\alpha_{CO}$ should also be affected by the broadened line widths (e.g., \citealt{shetty}). For a fixed column density, the optical depth of CO decreases as the velocity dispersion of the medium increases. As a result, the conversion factor for a fixed $\Sigma_c$ should be reduced by an amount $\approx\kappa R_c/\sigma_{sg}$ due to line broadening. 

This broadening of the line width by the galaxy potential has been invoked to explain the lower $\alpha_{\rm CO}$ in ULIRGs and galaxy centers (e.g., \citealt{downes},\citealt{bolatto2013}). Our model gives a way to treat galaxy disks and galaxy centers in a unified way. 

This also makes it possible to quantitatively unify apparently disparate virial mass and dust-based determinations of $\alpha_{CO}$ \citep{sandstrom,donovan}. In some cases, frequently targets with strong stellar bars, virial-based $\alpha_{CO}$ estimates remain high throughout the disk. Dust-based estimates, which do not depend on the line width, show lower values and central depressions in $\alpha_{CO}$ \citep{sandstrom}. 

Our model predicts this behavior based on broadening of the line by the galaxy potential. Figure \ref{fig:xco} shows an example calculation for the case of NGC 6946. We assume a weak bar potential and the rotation curve implied by this galaxy's stellar mass $M_{\star}=10^{10.5}M_{\odot}$ \citep{leroy08}. We adopt the typical size and surface density of clouds from \citet{donovan}. Our model predicts a low value for $\alpha_{\rm CO}$ (red curve) and suggests that linewidth-based approaches will yield higher values of $\alpha_{CO}$ than dust-based approaches.

\begin{figure}[t]
\vspace*{-.15in}
\includegraphics[width=1.\linewidth]{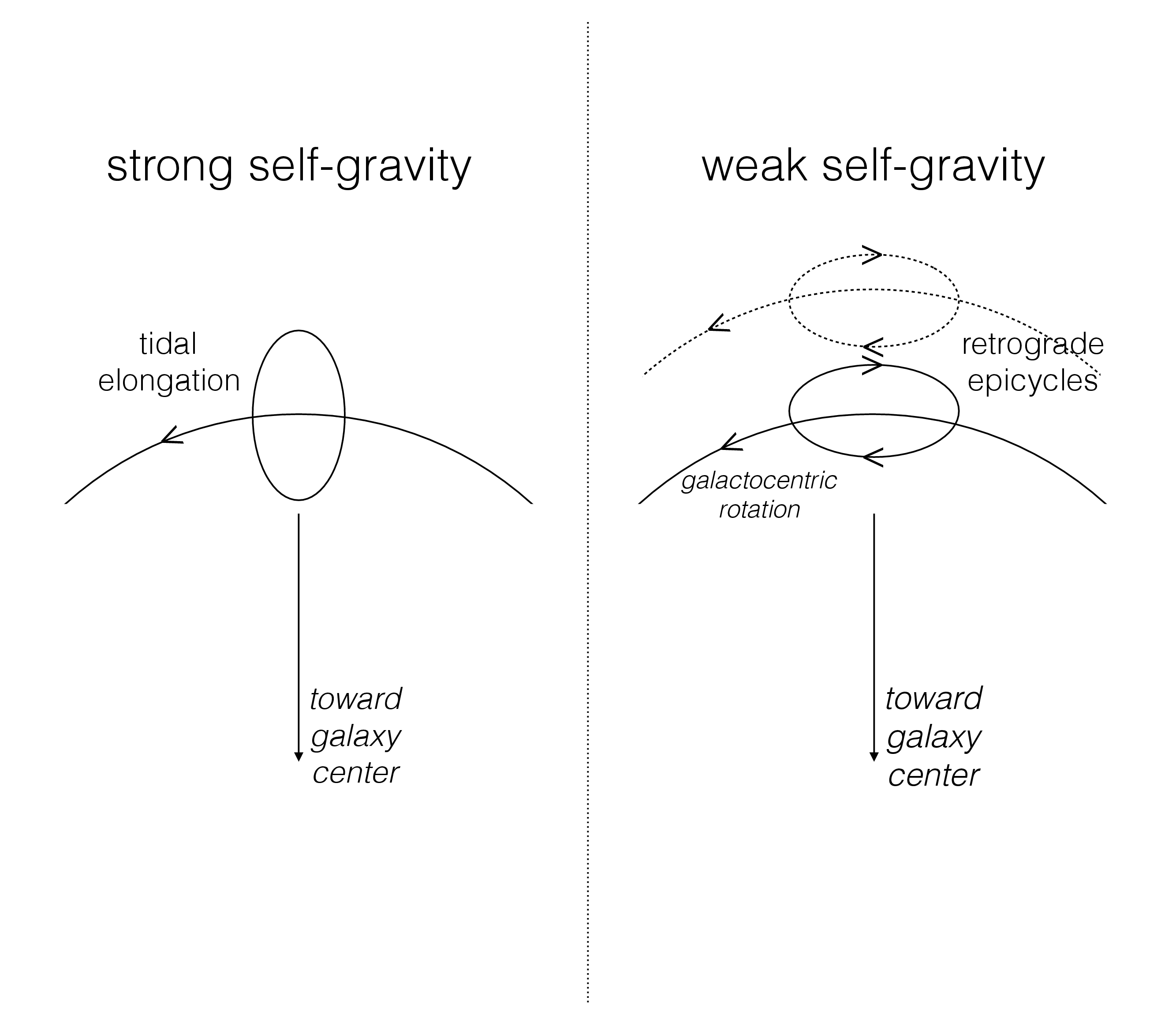}
\vspace*{-.15in}\caption{Illustration of cloud morphologies expected in the presence of the host galaxy potential in two regimes: (left) relatively strong self-gravity with negligible kinematic response to the galaxy potential and (right) relatively weak self-gravity and net kinematics following the epicyclic motions governed by the local galaxy potential.  On the left, the Coriolis force is negligible and the galactic tidal field lengthens clouds in the direction parallel to the gradient in the potential.  But on the right, the strong Coriolis force introduces motions along ellipses that are lengthened in the direction perpendicular to the potential gradient.  In this illustration the largest potential gradient is in the direction of the galaxy center, but another possibility would be a gradient transverse to spiral arms.  
\vspace*{0.35cm}  }
\label{fig:morphs}
\end{figure}

\subsection{Implications for cloud morphology}\label{sec:morphs}

Based on the previous section we identify two regimes in which cloud kinematics and morphology are significantly different from each other:  1) self-gravity is so high that internal ordered motions are relatively unimportant and the tidal force dominates over the Coriolis force, and 2) the local galactic potential dominates over self-gravity so that epicyclic motions exceed $\sigma_{sg}$. 

In the first, familiar case, clouds are expected to be elongated by tides in the direction of the galactic center. More generally, their long axis should point in direction of the strongest gradient in the potential.  (In the case of an axisymmetric potential, the strongest gradient in the potential is in the radial direction.)

In the second case, we would expect cloud morphologies to resemble the ellipse generated by epicyclic motion. These are elongated perpendicular to the largest gradient in the gravitational potential.  Although additional factors like cloud-cloud collisions and SNe feedback also have the potential to shape the morphologies of clouds, this suggests the interesting possibility that, under certain conditions, cloud morphology can serve as a measure of which  gravitational force (from self-gravity or the local galaxy potential) dominates. 

In the few extragalactic cloud surveys with sufficient resolution to assess cloud morphologies, cloud orientations are found to be diverse (i.e. \citealt{ros03}; \citealt{koda}; \citealt{hughesLMC}). This may imply a spectrum bounded by the two cases suggested here. High spatial (and spectral) resolution across a greater variety of environments would help improve tests of the model.

Interestingly, in M51, clouds in the two spiral arms are oriented with their long axes in the direction of the spiral (D. Colombo, private comm.). This is manifestly different from the orientation that would be expected in the presence of only the strong spiral arm tidal field, which would lengthen clouds across the arms, in the transverse direction.  According to our model, the perpendicular orientation arises when the galaxy imposes not only a tidal field but also impacts the internal motions within the cloud, i.e. when the cloud crossing time is comparable to the local orbital period (as discussed above).  

\section{Summary \& Conclusions}\label{sec:summary}

We consider the influence of host galaxies on their molecular clouds, exploring how cloud-scale gas motions reflect the interplay of the background galaxy potential and gas self-gravity. We suggest a picture in which clouds are not decoupled from the surrounding medium but rather influenced by the background galaxy and show that the model appears to agree with recent observations.

First, we formulate a new way of assessing the force exerted by the galaxy potential. We propose that the Coriolis force is important on the cloud scales when the crossing time of the cloud is comparable to the local galactic orbital period. In this case, the gas kinematics resemble motions in the limit of no self-gravity, where conservation of angular momentum restricts the gas to ordered, epicyclic motions.  We argue that an analogous case may hold in the vertical direction. 

The characteristic velocities of the epicyclic motions, expressed as a velocity dispersion, provide a straightforward measure of the strength of the host galaxy gravitational potential. This can be compared to other cloud-scale forces and (turbulent) motions. The epicyclic frequencies in the plane and in the vertical direction, which depend on the gradient of the gravitational potential, can be measured directly from the galaxy rotational velocity or inferred from the vertical mass distribution.  

For normal disk galaxies, observed rotation curves and galaxy shapes imply that epicyclic motions at the cloud scale can be comparable to the motions needed to support clouds against their own weight. In galaxy centers and in dynamical features like spiral arms, we find that galaxy-induced motions can even exceed motions due to self-gravity. Our model thus predicts an the important role for the galaxy potential in setting observed molecular cloud properties.  

In a preliminary comparison of the model to observations, we show that the model reproduces the observed line width at multiple spatial scales in the disk M51. It also reproduces observed deviations in cloud properties from those predicted in the purely self-gravitating case in a diverse population of clouds from a varied set of nearby galaxies.  

We discuss several natural next observational tests of this scenario. First, these coherent motions should be observable as velocity gradients on the scales of individual molecular clouds. These gradients are expected to show an alignment with the local bulk velocity field that depends on the intrinsic cloud shape.  Second, cloud morphologies might provide a useful probe of the relative dominance of self-gravity or the background potential. 

In the scenario we envision, the motions due to collapse under the influence of self-gravity add constructively to coherent motions induced by the background potential. This differs from the scenario in which gas motions reflect some form of self regulation, for example due to star formation feedback. Our initial tests support this idea, suggesting that super-virial motions correlate with the local potential. More tests and numerical simulations can both help clarify this crucial point.

Further tests of the influence of galactic environment on the ISM conditions that regulate star formation should be possible in the near future, with the growing number of high spatial and spectral resolution extragalactic molecular gas surveys currently in progress (\citealt{schinnerer}; Leroy et al.  in prep).  With maps of the star-forming reservoir at cloud-scale resolution that extend across a variety of galactic environments, we can place clouds in the context of their local galactic surroundings systematically for the first time. \\

Many thanks to the referee, Enrique V\'azquez-Semadeni, for a thorough review and a considered set of suggestions that improved the manuscript.  SEM acknowledges funding from the Deutsche Forschungsgemeinschaft (DFG) vsia grant SCHI 536/7-2 as part of the priority program SPP 1573 "ISM-SPP: Physics of the Interstellar Medium."  ER acknowledges the support of the Natural Sciences and Engineering Research Council of Canada (NSERC), funding reference number RGPIN-2017-03987.  JMDK and MC gratefully acknowledge funding from the German Research Foundation (DFG) in the form of an Emmy Noether Research Group (grant number KR4801/1-1, PI Kruijssen). JMDK acknowledges support from the European Research Council (ERC) under the European Union's Horizon 2020 research and innovation programme via the ERC Starting Grant MUSTANG (grant agreement number 714907, PI Kruijssen).  ES acknowledges funding from the European Research Council (ERC) under the European Union's Horizon 2020 research and innovation programme (grant agreement No. 694343).  The work of JP was partly supported by the Programme National "Physique et Chimie du Milieu Interstellaire" (PCMI) of CNRS/INSU with INC/INP co-funded by CEA and CNES.  FB acknowledges funding from the European Union's Horizon 2020 research and innovation programme (grant agreement No 726384 - EMPIRE).  AH acknowledges support from the Centre National d'Etudes Spatiales (CNES).  AU acknowledges support from the Spanish MINECO grants AYA2016-79006-P and ESP2015-68964-P.

\appendix
\begin{figure*}
\begin{tabular}{cc}
\hspace*{-.15in}\includegraphics[width=0.5\linewidth]{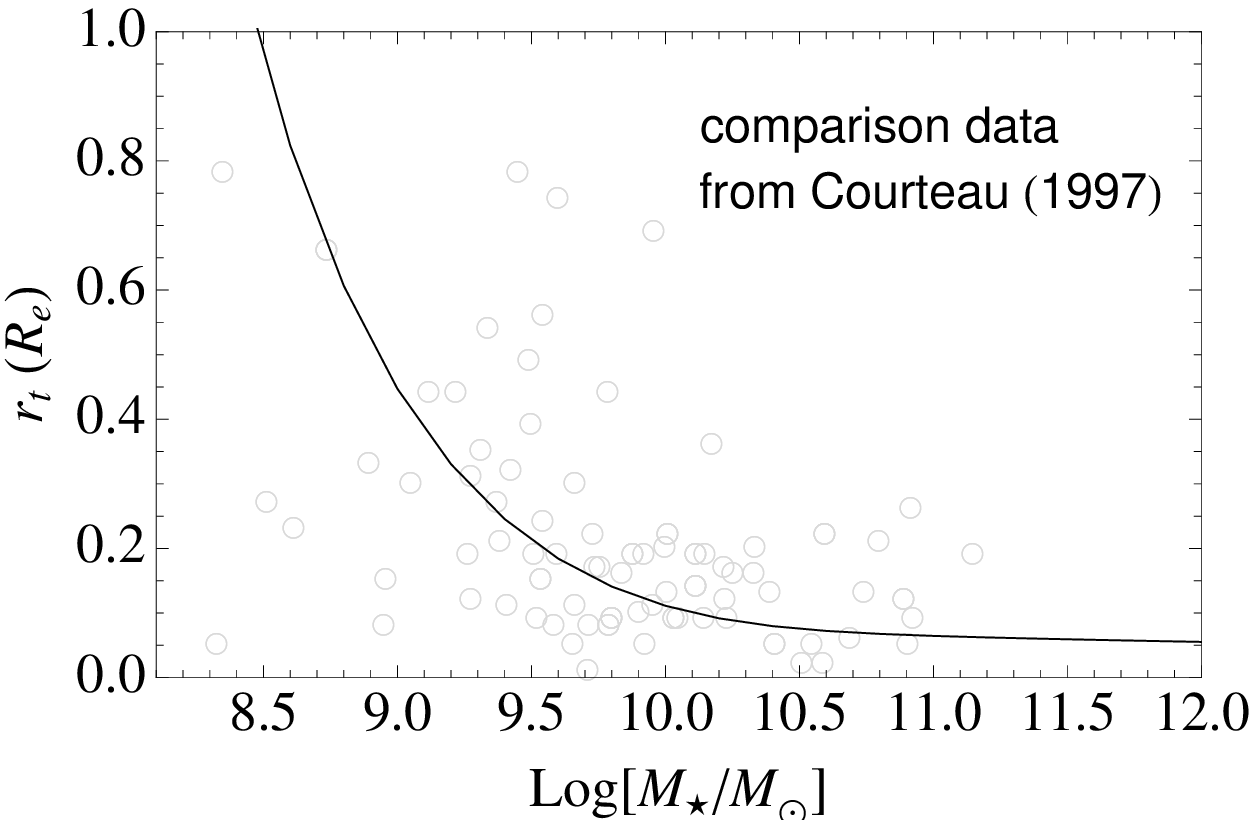}&\includegraphics[width=0.5\linewidth]{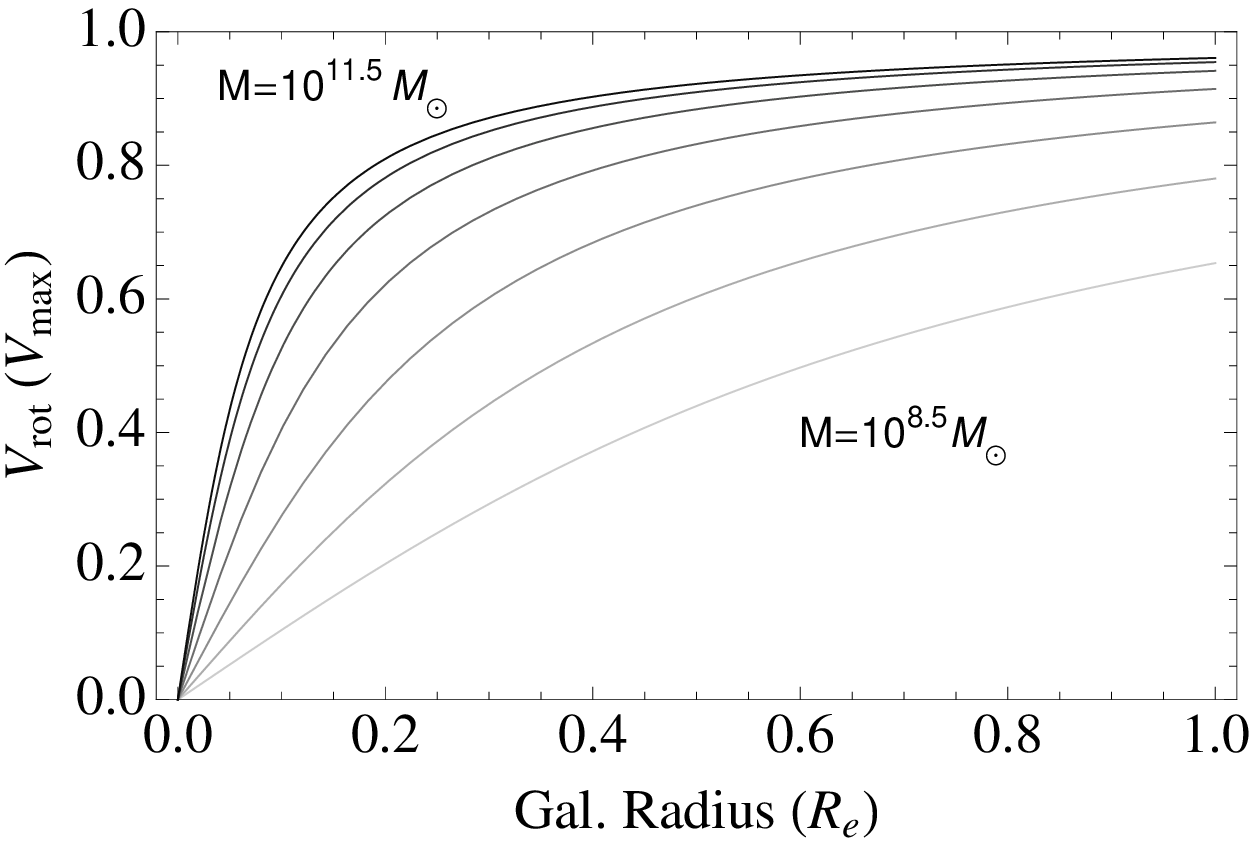}\\
\end{tabular}
\caption{(Left) The relation between the transition radius $R_t$ in eq. \ref{eq:rotationCurve}, parameterizing rotation curve shape, and galaxy stellar mass implied by the empirical relation between central stellar surface brightness (density) and inner rotation curve gradient measured by \cite{EF}, and assuming the scaling relations compiled by \cite{dutton1}. (Right) Galaxy rotation curve shape over a range of stellar masses, adopting the $R_t$ shown on the left, according to our empirically-based model (see text). Rotation curves of low mass galaxies rise much more slowly than those in high mass galaxies.  }
\label{fig:rotcurves}
\end{figure*}
\begin{figure*}[t]
\begin{tabular}{cc}
\includegraphics[width=0.5\linewidth]{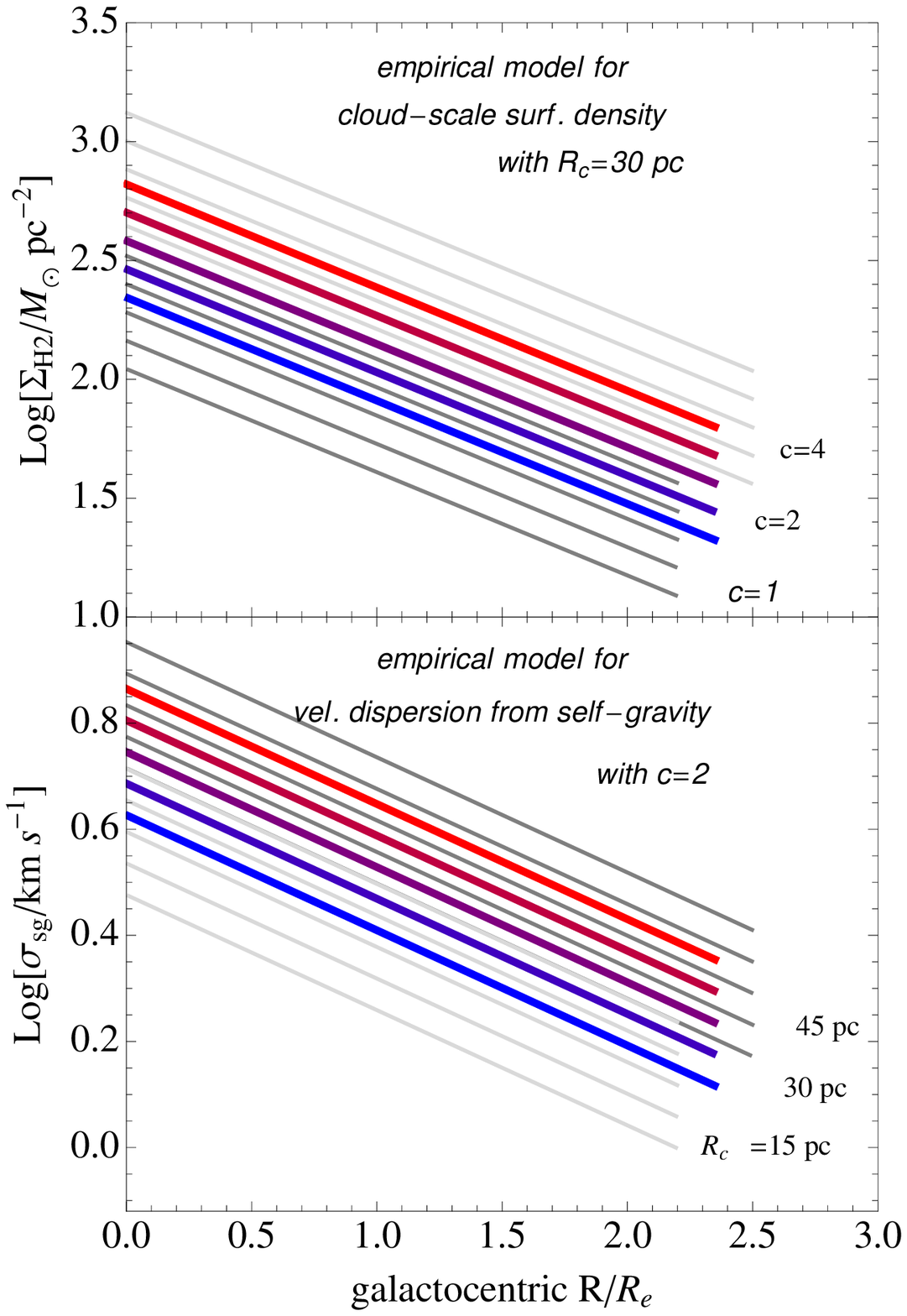}&\includegraphics[width=0.5\linewidth]{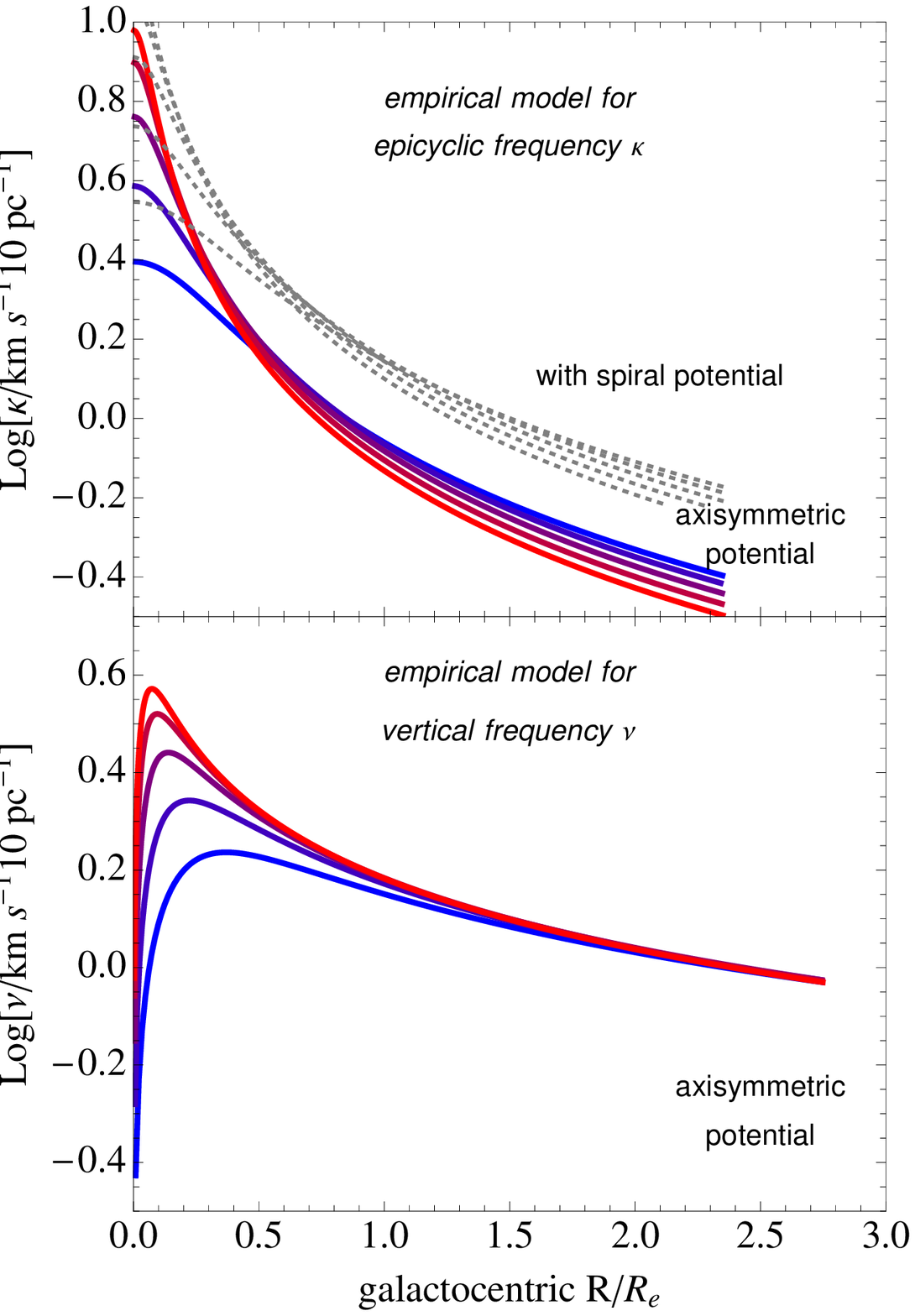}
\end{tabular}
\caption{(Top left) Empirical model for cloud-scale molecular gas surface densities based on global galaxy scaling relations (see text) and an assumed exponential distribution adjusted to the cloud-scale for three assumed clumping factors.  Five sets of profiles are shown for each $c$, one for each stellar mass in the range 9.25$<$$\log M_{\star}/M_\odot$$<$10.75 in steps of 0.5 $\log M_{\star}/M_\odot$. For the nominal $c$=2, lines are color-coded by galaxy stellar mass (from blue to red, low to high) and span the observed range of surface densities at 60pc cloud-scales in \cite{leroy2016}.  (Bottom left) Velocity dispersion balancing self-gravity in clouds with surface densities as shown in the top panel, specifically with the nominal $c$=2.  Sets of five profiles assuming three different representative cloud sizes are shown.  For the nominal $R_c$=30 pc lines are color-coded by galaxy stellar mass (from blue to red, low to high).  (Top right) Radial profiles of the epicyclic frequency $\kappa$ in axisymmetric disks (colored lines) with rotation velocities specified by our empirically-based rotation curve model.  Each line shows the profile implied at a stellar mass in the range 9.25$<$$\log M_{\star}/M_\odot$$<$10.75 in steps of 0.5 $\log{ M_{\star}/M_\odot}$.  The dotted gray lines show the radial variation of the local epicyclic frequency $K$ predicted in the presence of spiral density perturbations in the disk in eq. (\ref{eq:kappaspiral}).  (Top right) Radial profiles of the vertical epicyclic frequency $\nu$ in axisymmetric disks (colored lines) estimated according to eq. (\ref{eq:nuapprox}) adopting the stellar scale height implied at the stellar mass of the galaxy model.  
 }
\label{fig:empModelFig}
\end{figure*}
\begin{figure*}[t]
\begin{center}
\begin{tabular}{cc}
\includegraphics[width=0.45\linewidth]{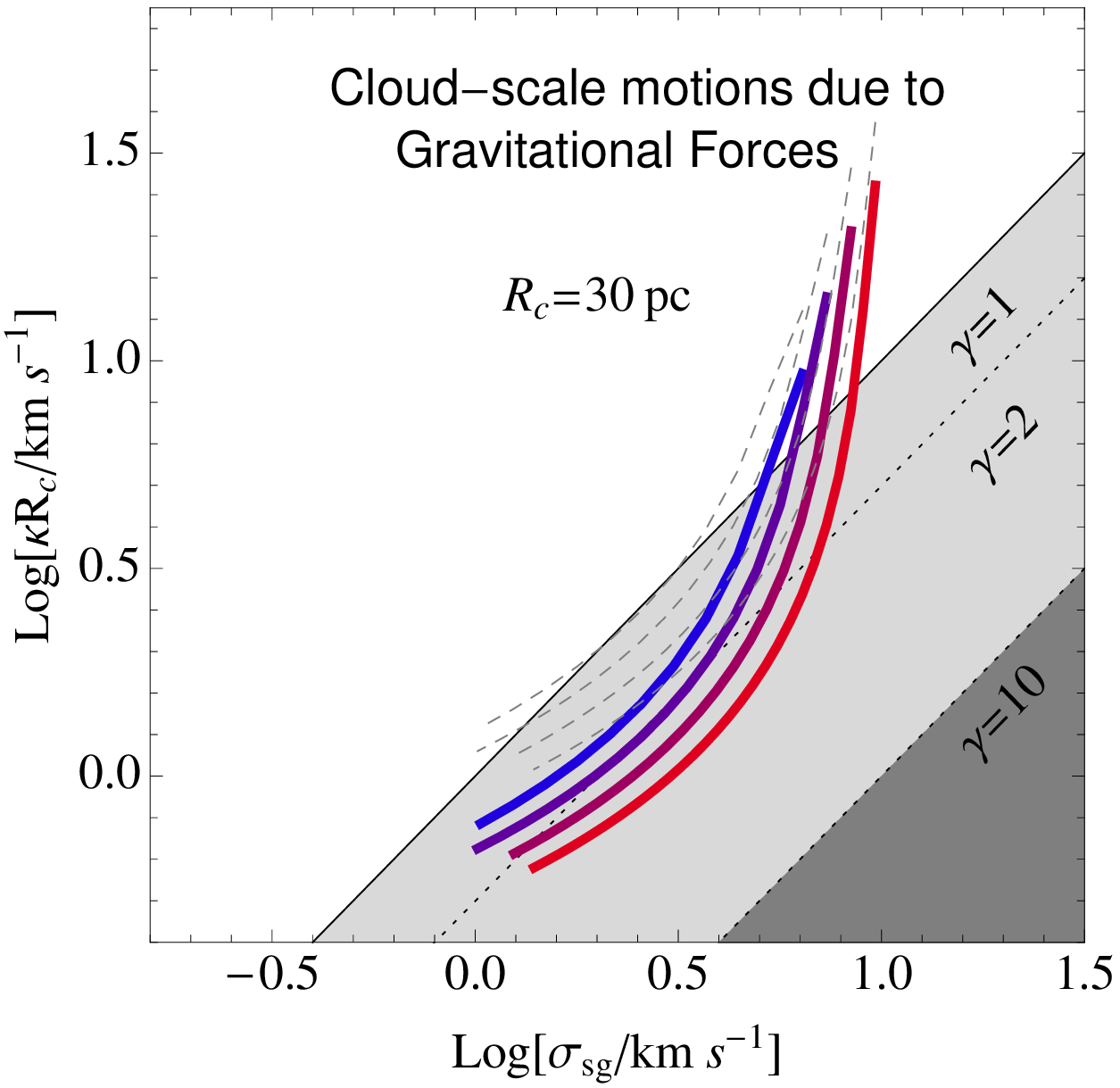}&\includegraphics[width=0.45\linewidth]{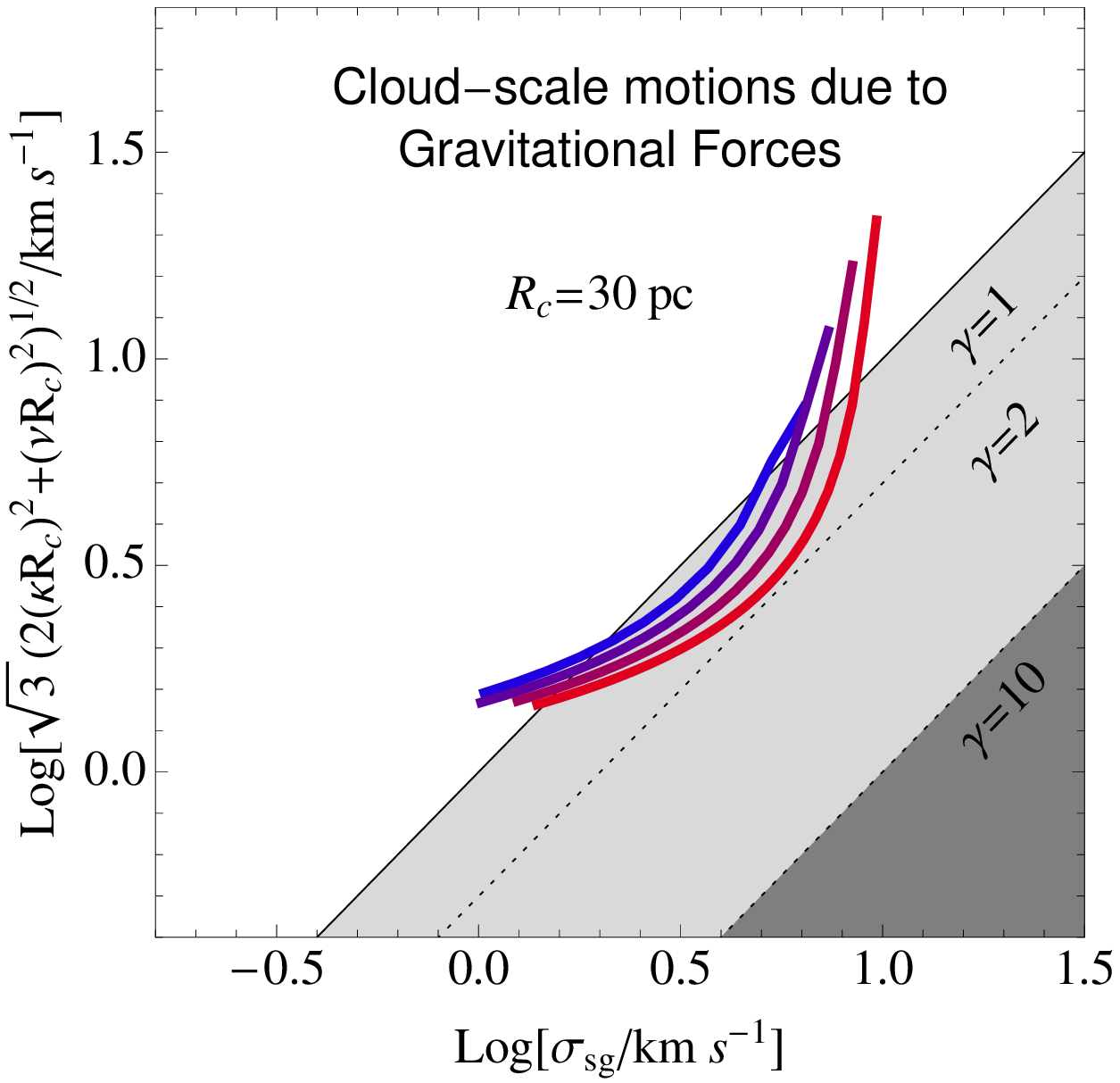}
\end{tabular}
\end{center}
\caption{Comparison of the relative strengths of gravitational forces on cloud scales.  The galaxy gravitational potential is represented by epicyclic motions in either the isotropic (left) or non-isotropic (right) cases on the vertical axis.  On the horizontal axis the motions $\sigma_{sg}$ due to the cloud's self-gravity are shown.   In each panel, a series of five curves illustrate the behavior of gravitational motions across the disks of five galaxies with stellar masses in the range 9.25$<$$\log M_{\star}/M_\odot$$<$10.75 in steps of 0.5 $\log M_\odot$, from left to right: low mass (blue) to high mass (red).  Curves span from the galaxy center (top) out to 4 disk scalelengths (bottom).  In the top right panel, the additional set of five dashed lines show the predictions at each stellar mass for the epicyclic frequency $K$ according to eq. (\ref{eq:kappaspiralEst}) illustrating the trends expected in the presence of a spiral perturbation to the galaxy potential.  Three diagonal lines highlight regimes of different relative force strength $\gamma_{sg}$=$\sigma/(\kappa R_c)$$\approx$$Q_c^{-1}$.  When $\gamma_{sg}$=1, gravitational forces balance.  When $\gamma_{sg}$$>>$1, the gas is self-gravitating.   \vspace*{0.35cm} 
 }
\label{fig:paramspace}
\end{figure*}

\section{A. A model for galaxy rotation curve shapes}\label{sec:appendix}
To assess the behavior of the epicyclic frequency $\kappa$ within galaxies, and from galaxy to galaxy, here we assemble a model for the shapes of rotation curves that depends on galaxy stellar mass.  Our model uses well-known scaling relations between galaxy size, mass and maximum circular velocity (as parameterized by the Tully-Fisher relation and the size-mass relation), in combination with empirical relations linking the gradient in the inner rotation curve gradient to stellar mass surface density.   

We assume a rotation curve of the form
\begin{equation}
V_{circ}=\frac{2 V_{max}}{\pi}\arctan{(R/R_t)} \label{eq:rotationCurve}
\end{equation}
where $V_{max}$ is the maximum circular velocity and $R_t$ is the so-called transition radius between the rising and the flat part of the rotation curve.  This form provides a good match to observed rotation curves for galaxies with a range of masses (i.e. \citealt{courteau}).  

For a galaxy of stellar mass $M$, we assign the scale length $R_e$ and $V_{max}$ using the empirical scaling relations in the local universe assembled by \cite{dutton1}.  
The relation between  $V_{max}$ and $M$ is based on the Tully-Fisher relation measured by \cite{bell} using their prescribed stellar M/L to convert K-band luminosity to stellar mass.   The scatter about the $V_{max}$-$M$ relation is quite small, and we translate this into a 1$\sigma$ uncertainty of 0.15 dex on $V_{max}$ at fixed $M$.  As measured by Shen et al. 2003, the local galaxy size-mass relation has an uncertainty of 0.2 dex on $R_e$ at fixed $M$.   

The shape of the rotation curve, as parameterized by $R_t$, is assigned at each $M$ using the observed correlation between the gradient in the inner rotation curve (measured inside $R_t$) and the bulge-to-total (mass) ratio found by \cite{EF}.  This strong correlation, which appears equally strong between the inner gradient and central stellar surface brightness, is testament to the sensitivity of the shape of the gravitational potential to the inner distribution of stellar mass.  

We estimate the bulge-to-total ratio at fixed stellar mass by adopting the relation between bulge and disk K-band luminosities appropriate for (late-type) Sab-Sac galaxies as measured by \cite{laurikainen}, assuming a constant mass-to-light ratio.  At these long wavelengths, a single M/L should be approximately equally valid for old, metal rich bulges and for young disks \citep{meidtmass}.  

From the bulge-to-total ratio at a given mass, which specifies the inner rotation curve gradient, we estimate $R_t$ given that the gradient in the rotation curve in eq. (\ref{eq:rotationCurve}) is 
\begin{equation}
\frac{dV_{circ}}{dR}=\frac{V_{max}}{\pi\left(1+\frac{R^2}{R_t^2}\right)R_t} \label{eq:gradient}
\end{equation}
such that at $R_t$ 
\begin{equation}
\left.\frac{dV_{circ}}{dR}\right\vert_{R_t}=\frac{V_{max}}{\pi R_t} \label{eq:gradientrt}
\end{equation}

The left panel of Figure \ref{fig:rotcurves} presents the $R_t$ implied by this empirically-based model as a function of galaxy stellar mass.  Shown for comparison are data from \cite{courteau} based on fits of eq. (\ref{eq:rotationCurve}) to observed rotation curves.   Our model provides a good representation of the completely independent Courteau measurements, notably capturing the lack of galaxies with large $R_t$ at masses $M_{star}$$>10^{10} M_{\odot}$.  The model is thus generally representative of real galaxies and parameterizes the expected slow rise in the rotation curves of low mass galaxies as compared to more massive galaxies, given that $R_t$ is a larger fraction of one disk scalelength $R_e$.  
This is illustrated in the right panel of Figure \ref{fig:rotcurves} showing the (normalized) rotation velocity as a function of galactocentric radius (in terms of $R_e$) for galaxies in the mass range $10^{8.5}$$<$$M_{star}/M_{\odot}$$<$$10^{11.5}$.  

\section{B. A model for vertical gas motions in disk galaxies}\label{sec:appendixnu}
The frequency of vertical epicyclic motions $\nu$ in the molecular gas in star-forming disk galaxies is set by the vertical shape of the background disk gravitational potential defined by the stars and dark matter.  To estimate $\nu$ we use eq. (\ref{eq:nuapprox}) together with an empirically-based model for $\kappa$ and an estimate of the stellar scale height $z_0$ at a given stellar mass.  
Observations of edge-on late-type disk galaxies analyzed by \cite{bershady} suggest 
\begin{equation}
(z_0/kpc)\approx0.2 (R_e/kpc)^{0.68} 
\end{equation}
to within 0.1 dex, where $R_e$ is the scale length of the stellar disk that we link to stellar mass through the empirical galaxy size-mass relation used in Appendix \ref{sec:appendix}.  Our nominal model assumes that a constant scale height applies at all galactocentric radii.  The radial variation in $\nu$ estimated according to this model is shown in Figure \ref{fig:empModelFig}, together with the epicyclic motions estimated from the rotation curve predicted at a given galaxy stellar mass (see Appendix \ref{sec:appendix}).  Note that this model, which is intended to describe motions in the centralized, molecule-rich portions of galaxy disks, does not take into account the increasing dark matter contribution to the background disk density with increasing radius.  This would increase $\nu$ above the value shown in Figure \ref{fig:empModelFig} moving towards galaxy outskirts.  In contrast, a flared disk, in which the scale height increases with radius, would reduce $\nu$ progressively more with radius.  

\section{C. A model for cloud-scale molecular gas surface densities in normal star-forming galaxies}\label{sec:appendixGas}

In order to estimate the motions $\sigma_{sg}$ due to gas self-gravity, here we develop a simple model for the gas surface density on cloud scales expected in star-forming disks.  From the parameterization of the Star formation main sequence measured by \cite{chang} in the local universe we first assign the SFR at fixed stellar mass.  Then we use the observed linear relation between molecular gas and SFR surface densities in nearby galaxies (specifically the relation measured by Leroy et al. 2013 corresponding to a molecular gas depletion time $t_{dep}$=$\Sigma_{mol}/\Sigma_{SFR}$=2 Gyr) to estimate the molecular gas surface density $\Sigma_{H_2}$.  For this conversion we assume that the distributions of gas and young stars are identical (with the same radial scalelengths) and that the gas scale length $R_e^{gas}$=0.83$R_e^{*}$ \citep{bolatto2015}.  An additional clumping factor $c$ lets us estimate the surface density on cloud scales $\Sigma_c=c\Sigma_{H_2}$.  \cite{leroy2013b} show that $c$ ranges between 1-10 in molecular gas.  Here we choose $c$ so that our model for $\Sigma_{H_2}$ can match the surface densities measured by \cite{leroy2016} on 60 pc scales, probing characteristic cloud sizes $R_c$$\approx$ 30 pc.  With a nominal value  $c$=2 our estimated $\Sigma_{c}$ inside $1R_e$  (see top left panel of Figure \ref{fig:empModelFig}) ranges on average from 70 $M_\odot pc^{-2}$ to 500 $M_\odot pc^{-2}$ across the galaxy stellar mass range $10^{8.5}$$<$$M/M_{\odot}$$<$$10^{11.5}$.  To reach these surface densities assuming a depletion time shorter by a factor of 2 requires doubling the clumping factor.  

\subsection{Average trends in gravitational motions within galaxies with a range of masses}\label{sec:quantplots2}
Figure \ref{fig:paramspace} demonstrates how $\sigma_{sg}$ and $\sigma_{gal}$ compare on average in galaxies with a range of galaxy stellar masses, using the simple empirically-motived model for the average gas surface density described above to estimate $\sigma_{sg}$ at each stellar mass using eq. (\ref{eq:sigSG}).  For each galaxy mass we adopt the rotation curve model developed in  Appendix \ref{sec:appendix}, which provides a measure of $\kappa$ (according to eq. \ref{eq:nuapprox}) and $\nu$ following Appendix \ref{sec:appendixnu}.  

The extent of the curves in Figure \ref{fig:paramspace}, which each follow the radial variation of both sets of motions in a given galaxy disk, is intended to capture the full range out to the edge of the molecule-bright emission.  In terms of the $H_2$ scalelength $R_e$=0.2 $R_{25}$  measured in a sample of nearby galaxies by \cite{schruba}, typically half of the total CO flux tracing molecular hydrogen is enclosed within a radius $R_{50}\approx$1.5 $R_e$ (very near the transition from HI to H$_2$) and 90\% is within $R_{90}\approx$ 4$R_e$  \citep{schruba}.  

Both quantities $\sigma_{sg}$ and $\kappa R_c$ in Figure \ref{fig:paramspace} are shown at a chosen scale of $R_c$=30 pc, which is typical of measured cloud sizes in the MW (\citealt{heyer}; \citealt{mv17}) and external galaxies (e.g. \citealt{bolatto}; \citealt{hughesI}; \citealt{leroy2015}). 
Realistic variations in cloud size and gas clumpiness (not illustrated) are insufficient to drastically alter the position of the predicted curves in this parameter space.  Only with extreme parameter choices do the predictions approach our conservative estimate for the onset of self-gravity at $\gamma_{sg}=Q^{-1}$=10 in the figure. (Our empirical calibration suggests a value closer to $\gamma_{sg}$=12.)   

The locus of models in this parameter space demonstrates that the gravitational forces due to cloud self-gravity and the local galaxy potential are comparable on cloud scales.  As also described in $\S$ \ref{sec:quant}, galaxy-induced motions can even exceed those associated with the cloud's self-gravity in the centers of galaxies and towards the edge of the molecular disk, with a more rapid increase with radius in the non-isotropic case than in the more conservative isotropic case.  

As highlighted by the gray dashed lines, besides galaxy centers, stellar dynamical features like spiral arms can be locations where epicyclic motions become increasingly important relative to $\sigma_{sg}$.  As modeled in $\S$ \ref{sec:spiralkappa}, spiral arm perturbations introduce locally large gradients in the gravitational potential, raising the epicyclic frequency (eq. \ref{eq:kappaspiral}).  In these environments, the background host galaxy makes an increased contribution to observed motions, which are thus raised above the level expected due to only gas self-gravity.

\end{document}